# Probing the interplay between lattice dynamics and short-range magnetic correlations in CuGeO$_3$ with femtosecond RIXS

E. Paris[1,†,*], C. W. Nicholson[2,†], S. Johnston[3], Y. Tseng[1], M. Rumo[2], G. Coslovich[4], S. Zohar[4], M.F. Lin[4], V.N. Strocov[1], R. Saint-Martin[5], A. Revcolevschi[5], A. Kemper[6], W. Schlotter[4], G. L. Dakovski[4], C. Monney[2], and T. Schmitt[1,*]

[1]*Photon Science Division, Paul Scherrer Institut, CH-5232 Villigen PSI, Switzerland*

[2]*Départment de Physique and Fribourg Centre for Nanomaterials, University of Fribourg, CH-1700 Fribourg, Switzerland*

[3]*Department of Physics and Astronomy, University of Tennessee, Knoxville, TN 37996 USA*

[4]*Linac Coherent Light Source, SLAC National Accelerator Laboratory, Menlo Park, CA 94025, USA*

[5]*Laboratoire de Physico-Chimie de l'Etat Solide, ICMMO, Université Paris-Saclay, 91405 Orsay Cedex, France*

[6]*Department of Physics, North Carolina State University, Raleigh, NC 27695, USA*

†Equally contributing authors

*Corresponding authors. Email: <u>eugenio.paris@psi.ch</u> (E.P.); thorsten.schmitt@psi.ch (T.S.)

## Abstract

Investigations of magnetically ordered phases on the femtosecond timescale have provided significant insights into the influence of charge and lattice degrees of freedom on the magnetic sub-system. However, short-range magnetic correlations occurring in the absence of long-range order, for example in spin-frustrated systems, are inaccessible to many ultrafast techniques. Here, we show how time-resolved resonant inelastic X-ray scattering (trRIXS) is capable of probing such short-ranged magnetic dynamics in a charge-transfer insulator through the detection of a Zhang-Rice singlet exciton. Utilizing trRIXS measurements at the O *K*-edge, and in combination with model calculations, we probe the short-range spin-correlations in the frustrated spin chain material CuGeO$_3$ following photo-excitation, revealing a strong coupling between the local lattice and spin sub-systems.

## Introduction

Among the family of transition metal oxides, the charge-transfer materials are highly studied due to the realization of a number of exotic properties, from metal-insulator transition to high-Tc superconductivity. These phenomena often arise from the microscopic coupling between charge, spin, orbital and lattice degrees of freedom. In this regard, a significant advancement in understanding the origin of these exotic properties may be obtained by employing ultrafast time-resolved techniques to probe the dynamics of the relevant ordered phases. In the study of magnetism, the interplay between magnetic sub-lattices[1] and other degrees of freedom such as the crystal lattice[2–5] and charge[6–10], have afforded numerous important insights. However, when materials crystallize in a purely one-dimensional (1D) crystal structure they cannot support long-range magnetic order[11], even in the presence of strong short-range magnetic correlations. Such correlations are not easily accessible to many ultrafast techniques such as time-resolved photoemission, x-ray diffraction, or optical spectroscopies. In contrast, Resonant Inelastic X-ray Scattering (RIXS) is able to probe both local magnetism and access elementary excitations rather than ordering phenomena[12–14]. Moreover, in contrast to most other local magnetic probes, the intrinsic timescale of the RIXS process (~1 fs) easily allows the extension of this technique to the ultrafast domain[15]. The main limitation for this class of experiments lies in the limited available time-integrated intensity. Indeed, only recently the progress of soft x-ray free-electron lasers (FELs) has allowed the very first few experiments of this kind [16–20]. The advent of the next-generation FELs worldwide is expected to overcome these limitations, enabling the study of a wide class of problems in condensed matter and beyond.

Therefore, using trRIXS holds great promise for widening our understanding of ultrafast magnetism in low-dimensional and frustrated magnetic systems, allowing insights into their rich physics that include spin glass phases, novel types of elementary excitations and the fractionalization of quasi-particles. In this class of materials, the complexity of the phase diagrams often results from the competition between nearly degenerate ground states, and a close competition between nearest neighbor (NN) and next-nearest neighbor (NNN) magnetic exchange coupling. A typical example of such physics is realized in the material $CuGeO_3$, whose structure is shown schematically in Fig. 1(a). The basic building blocks of $CuGeO_3$ are $CuO_4$ plaquettes arranged in edge-shared chains running along the crystallographic $c$-axis[21]. The exchange interaction, $J$, between NN Cu ions depends on the interatomic distance and is antiferromagnetic (AFM), in large part due to a Cu-O-Cu bond angle of 99°[22]. The system is unstable towards a lattice dimerization, opening a spin-Peierls (SP) gap in the magnetic spectrum below a temperature $T_{SP} = 14$ K[23–26]. This process involves a magneto-elastic coupling between the 1D electronic system and the 3D phonon system. While the classical description of a SP transition holds only when the phonon energy is small compared to the other relevant energy scales, for $CuGeO_3$ the phonon energy scale is of the same order as the NN exchange coupling $J$[27]. The resulting entanglement of spin and lattice degrees of freedom is further complicated by the magnetic frustration induced by the presence of a large NNN AF exchange. In further contrast to typical SP materials such as TiOCl where the lattice dimerization is associated with the softening of a particular phonon mode[28,29] no soft phonon was ever observed in $CuGeO_3$. Even more remarkably, the two modes most strongly associated with the distortion are found to harden at low temperatures[30]. An additional quasi-elastic mode appears due to short-range fluctuations already well above the transition temperature[31] as evident in a number of experiments[32–34]. Therefore, due to the concomitant effect of low dimensionality and geometrical frustration, $CuGeO_3$ does not order magnetically over a long-range, and the physics is instead dominated by short-range magnetic correlations.

To obtain a better understanding of the intricate relation between spin and lattice degrees of freedom in this and other low-dimensional oxide materials, a direct measurement of the dynamics of short-range spin correlations is highly desirable. Here, we use time-resolved RIXS (trRIXS) to shed light on the interplay between lattice modulations, initiated by an ultrashort photoexcitation, and short-range magnetic correlations in a frustrated magnetic system. We employ 4.7 eV photons to excite the system across the charge-transfer gap, and O $K$-edge RIXS to track the photo-induced response of a Zhang-Rice singlet (ZRS) exciton, which is sensitive to the short-range spin correlations within the Cu-O chain of $CuGeO_3$[35]. Following photoexcitation, we observe an exponential decrease of the ZRS intensity, with some deviation caused by a partial recovery within ~1 ps, followed by a gradual reduction on longer time scales (10 ps), which persists to long delays (500 ps). This long-time effect increases linearly with the pump fluence up to a saturation above a critical value of 5 mJ/cm$^2$, which suggests the removal of a coupling channel between the magnetic and lattice sub-systems. Exact diagonalization calculations of the RIXS spectra at delays < 10 ps imply the influence of a damped phonon which modulates the on-site Cu energy, therefore affecting the spin-spin correlation function. The timescale involved is compatible with a magneto-elastic mode at ~1 THz. In this way, we explore the effect of lattice dynamics on the short-range spin correlations in a frustrated low-dimensional spin system, and set the stage for future investigations of these interactions exploiting next generation x-ray FEL sources.

## Results

**Zhang-Rice singlet detection with O K-edge RIXS**. In our measurements, the energy of the FEL x-ray pulses is tuned to the O *K*-edge (~ 531 eV) and set to be resonant to an absorption peak sensitive to the upper Hubbard band (UHB) electrons[35]. The O K-edge XAS spectrum, collected with a synchrotron source, is presented in Fig. 1(b). Fig. 1(a) shows the schematics of the trRIXS experiment: the pump and probe pulses propagate collinearly and impinge with an angle of 45° on the sample surface. In the experimental geometry, the CuO$_4$ plaquettes of copper-oxygen chains of CuGeO$_3$ lie at an angle of 56° with respect to the cleavage plane. The sample is kept at a temperature of 20 K during the measurements. Fig. 1(c) displays a comparison between the O *K*-edge RIXS spectrum obtained in the trRIXS experiment (lower spectra) and a static high-resolution spectrum obtained under the same experimental conditions at a synchrotron source. Despite the lower statistics and energy resolution available in the FEL experiment, all of the main spectral components such as the quasi-elastic, d-d, and charge-transfer excitations are clearly visible. In particular, an excitation located at 3.8 eV[35] is well resolved and separated from the broad charge-transfer structure, located between 4 and 10 eV energy loss. Such excitation is due to the formation of a Zhang-Rice Singlet (ZRS) state.

In order to understand the relevance of the ZRS excitation for probing short-range spin correlations, we first clarify the mechanism of the ZRS formation in the RIXS experiment[35]. In the initial state (see Fig.1 (d)), the Cu ions in two neighboring CuO$_4$ plaquettes have a (3d$^9$, 3d$^9$) orbital configuration, with a single hole in a hybridized Cu 3d$_{x^2-y^2}$/O 2p orbital per plaquette. We assume an initial configuration with antiparallel spins. When an x-ray photon is absorbed, it promotes an electron from the oxygen 1s core level to fill a hole in the valence band and the plaquettes assume a (d$^9$, d$^{10}$1̲s̲) configuration. From this excited state, a possible de-excitation channel involves the relaxation of one ligand electron from the neighboring plaquette to fill the core-hole, leaving the system in a (d$^9$L̲, d$^{10}$) configuration (see Fig. 1(e)). This mechanism involves a nonlocal charge transfer process, which is detected by the RIXS measurement[35,36]. In the RIXS final state, the two holes residing on the same plaquette inherit the initial antiparallel spin configuration and adopt the ZRS wave function. As a result of this bound ZRS exciton, the process gives rise to a peak at an energy transfer smaller than the charge gap (Δ) and separated from the continuum of charge-transfer excitations. It has been shown that the probability for such a process depends on the tendency of the neighboring Cu spins to be AFM oriented[35,37,38]. For this reason, the strength of the short-range AFM correlations is encoded in the spectral weight of the ZRS excitations probed by RIXS.

We excite the system across the charge transfer gap using a 4.7 eV ultraviolet laser pulse. By comparing the O *K*-edge RIXS spectrum collected before (-1 ps) and after (+6 ps) the photoexcitation, presented in Fig. 1(c), one can clearly identify a suppression in the intensity of the ZRS excitation as compared to the rest of the spectrum, which remains essentially unchanged.

**Dynamics of the short-range antiferromagnetic correlations**. Changing the pump-probe delay allows us to follow the development of the O K-edge trRIXS spectrum in the time domain. Although all the different excitations in the RIXS spectrum show some evolution following the arrival of the pump pulse, only the changes in the ZRS are clearly visible above the noise level. Therefore, in the following we focus our analysis on the evolution of the ZRS intensity as a function of the pump-probe delay. Figure 2(a) shows the O *K*-edge RIXS spectra between the elastic line and the charge transfer structure measured as a function of the time delay for a laser fluence $F = 37.4$ mJ/cm$^2$. Such a high incident fluence is used due to the large penetration depth of photons at the pump wavelength[39,40] and corresponds to an excitation of ~ 0.023 electrons per unit cell (see the Supplementary Materials, section S1, for the derivation of this value). The RIXS spectra in the region of the ZRS excitation are presented in Fig. 2(d) for selected time delays. After 0.5 ps from the arrival of the pump pulse, we observe a rapid suppression of the ZRS intensity and a partial recovery, as evidenced by a plateau in the dynamics ($\approx$ 1 ps). This is followed by a further suppression which does not recover to the original intensity even after >100 ps (see Fig. 3(d)). By employing a multi-component Gaussian fitting (as exemplified in Fig. 2(b)), we extract the dependence of the ZRS integrated intensity as a function of the time delay, as shown in Fig. 2(c), which confirms the dynamics already ascertained from the raw data. Directly after the laser excitation, we detect the fast suppression of the ZRS intensity with the additional plateau (~1 ps) followed by a longer reduction of intensity, saturating after ~10 ps. We note that the short time behavior depends strongly on the fitting model: in supplementary Fig. S6c this plateau instead appears as a distinct peak, which we discuss further below. While the dynamics on the longer timescale are probably dominated by the quasi-thermal heating of the lattice, the short timescale plateau may have a non-thermal origin. In the following, we will address first the slow dynamics and then return to the origin of the rapid non-thermal behavior.

The time-dependent behavior up to long time-delays is presented in Fig. 3 (d) for two different values of the laser fluence. The magnitude of the suppression of the ZRS peak intensity shows a large fluence-dependence, being prominent at F = 37.4 mJ/cm$^2$ and almost negligible at F = 2.5 mJ/cm$^2$. By fixing the time delay at 100 ps, we have investigated the ZRS intensity for various pump excitation fluences, as presented in Fig. 3(a). Using the fit procedure described before to extract the reduction of the ZRS reveals the onset of an abrupt saturation of the suppression for $F > 5$ mJ/cm$^2$ (see Fig. 3(b)). Since these data are acquired at long delays, we initially assume that they predominantly reflect the reduction of the ZRS intensity caused by the increase in the thermal energy of the lattice at quasi-equilibrium with the spin and electronic systems. In the 1D crystal structure, the neighboring chains are decoupled, which results in inefficient heat transfer away from the probed volume into the crystal and may explain the observation that, even at very long delays, the ZRS intensity remains strongly depleted for high incident fluences. The long time value of the ZRS intensity can be converted into an effective temperature by assuming a one-to-one correspondence with temperature-dependent static RIXS data (see Section S3 of the Supplementary Materials). The effective temperature, reported as a function of the fluence in Fig. 3(c), saturates at a value of $\approx$ 230 K. In contrast, the static temperature-dependent data only present a saturation behavior above room temperature. Such a deviation from the equilibrium behavior is surprising as heat transport is likely to dominate on such long time-scales, and at higher fluences more thermal energy is pumped into the system.

The same fitting analysis and comparison with static data can be applied to the transient ZRS behavior to extract the quasi-thermal evolution of the magnetic sub-system during the ultrafast measurements. The transient reduction of the ZRS peak as a function of time, again for $F = 37.4$ mJ/cm$^2$ [Fig. 2(c)], is fitted with an exponential decay (see Section S3 of the Supplementary Materials). In order to gain further insight into these dynamics we perform a commonly applied two-temperature model analysis. The model assumes that the electronic and lattice sub-systems are coupled heat baths that exchange energy following a pulsed excitation, which is captured by a system of two coupled differential equations (see Section S3 of the Supplementary Materials for full details). The evolution of the electronic ($T_{electronic}$) and lattice temperature ($T_{lattice}$) can therefore be estimated from the electronic and lattice specific heats, and the amount of energy deposited by the laser pulse. The temporal evolution of the effective magnetic temperature can be quantitatively reproduced by this simple model, as shown in Fig. 3(e), but only by assuming an absorbed heat content substantially lower than the amount estimated in the experiment. Indeed, by considering the mismatch in the penetration depth between the pump and the probe beams, for an incident laser fluence of 37.4 mJ/cm$^2$ at the sample surface, we estimate an absorbed fluence of 140 mJ/cm$^3$ within the volume probed by the x-rays (see Supplementary text S1). However, using the estimated 140 mJ/cm$^3$ as the input to the two-temperature model results in a dramatic overestimation of the rise in temperature when compared with that obtained from the ZRS intensity reduction. In fact, our model requires an input fluence of only 83 mJ/cm$^3$ to describe correctly the experimental data. Thus, the magnetic system is cooler than expected as compared with a simple two-temperature analysis of the electronic and lattice systems. This discrepancy further implies that the ZRS selectively probes the dynamics of the magnetic sub-system, and that this becomes decoupled from the lattice and electronic systems in accordance with the fluence dependent results discussed above.

**Non-thermal behavior at short timescales**. A possible scenario resulting in the plateau in the otherwise exponential decrease of the ZRS spectral weight dynamics is a partial recovery caused by a damped coherent phonon oscillation, modulating the local magnetic correlations. Indeed, a large involvement of the lattice degrees of freedom was found in the transient response of CuGeO$_3$ following photoexcitation using optical techniques[41,42]. Even at equilibrium, the magnetic and lattice degrees of freedom are known to be strongly coupled in CuGeO$_3$. In particular, the bond angle between the Cu-O-Cu atoms of the plaquettes sensitively affects the AFM coupling strength. Even before the transition to the SP-phase, the lattice undergoes structural changes by compressing the plaquettes along the *a*-axis and extending them along the *c*-axis[22]. This compression further increases the Cu-O-Cu bond angle and, therefore, enhances the AFM correlations. Below $T_{SP}$ an additional motion of the Ge side group is known to occur during the formation of the SP order[22,27,30], which also affects the AFM coupling.

In order to address whether a coherent phonon oscillation can lead to a modulation of the short-range magnetic order, we have utilized a cluster model of the plaquettes (Ref.[35] and Section S4 of the Supplementary Materials). To capture the short-time dynamics of CuGeO$_3$, we introduced a coherent oscillation in our cluster, coupling linearly and uniformly to the charge transfer energy $\Delta = \varepsilon_p - \varepsilon_d$. We also introduced an effective temperature in our model, to capture the long-time dynamics. The RIXS spectra are then calculated at each time delay using exact diagonalization and the Kramers-Heisenberg formalism (see further details in Section S4 of the Supplementary Materials). The results of this calculation in the region of the ZRS are presented in Fig. 4(a) with cuts at selected time delays shown in Fig. 4(b). Modulating the charge-transfer energy of the chains in the manner of a damped phonon oscillation, shown in Fig. 4(c) provides us with a possible explanation for the plateau at 1 ps in the observed fast dynamics of the ZRS intensity [Fig. 4(d)], namely the reduction and recovery after laser excitation. The slow reduction is also well reproduced by the effective temperature based on the experiment, as described above.

## Discussion

Our results reveal that the evolution of the magnetic degrees of freedom in $CuGeO_3$ decouple from the other degrees of freedom on a 100 ps time scale and possibly follow a non-thermal behavior on the 1 ps time scale.

The observation of a saturation of the ZRS intensity as a function of fluence at $\Delta t = 100$ ps (Fig. 3(b)) points towards the loss of an efficient coupling channel between the spin and phonon systems out-of-equilibrium. We note that while it is conceivable that this saturation could be caused by the suppression of the ZRS intensity due to the thermally heated lattice, such a scenario does not explain why the saturation occurs already at an equivalent temperature of ≈ 230 K, and not 300 K as observed in static RIXS (see Supplementary Materials, Fig. S7). This suggests a non-thermal behavior of the magnetic sub-system on a 100 ps timescale. Furthermore, it strongly implies the ZRS intensity is not a good measure of the quasi-equilibrium lattice temperature at all times, and can only reliably be used to probe the effective magnetic sub-system temperature. This conclusion is additionally supported by our two temperature model, which requires a significantly lower input energy density than in the experiment to reproduce the observed magnetic temperature.

A potential origin for this loss of coupling between the two sub-systems is the removal of a low energy magneto-elastic mode at 0.9 THz observed in the SP-phase[43–49]. It has been suggested that this mode is a bound pair of magnons held together by the spin-phonon interaction[50,51] giving it a strong phononic component. Since this mode is gradually removed by temperature at thermal equilibrium, it may also be quenched by the increased quasi-equilibrium temperature on the 10's of ps timescale.

Our model calculation reveals that a damped oscillation could be responsible for the observed non-thermal ultrafast dynamics of the ZRS. We emphasize that the plateau feature in our data depends strongly on the model used to fit the trRIXS spectra, and in particular whether the width of the ZRS is held constant or is left as a free fit parameter. In supplementary Fig. S6 (c), we show that, in the latter case, the plateau is replaced by a clear peak in the time dynamics, with the ZRS width also changing rapidly after excitation. This gives further weight to the hypothesis that the plateau obtained in the more conservative analysis is indeed the result of a damped oscillation. However, there remains the possibility that a non-equilibrium state occurs where the ZRS is broadened non-thermally. Given the currently limited statistics we leave this question open for future investigations. From the data in Fig. 2(c), we can estimate an energy scale of ~ 4 meV (1 ps) from the frequency of this mode. There remains the question of exactly which atomic motion causes the change of the magnetic coupling. For the NN exchange in $CuGeO_3$, there are two significant factors that lead to an overall AFM nature of the coupling. The first is the fact that the Cu-O-Cu bond angle differs significantly from 90°, which removes the symmetry restriction on superexchange imposed by a 90° bond. The second is that the degeneracy of the O $p_x$ and $p_y$ orbitals is removed by the presence of a Ge side-group out of the plane of the CuO$_4$ plaquettes[22,52], allowing additional AFM coupling. Both the Cu-O and the Ge-O bonds change at the SP-distortion, and both may contribute to the modulation of the charge transfer energy ($\Delta$), which equivalently modulates *J*. However, the frequencies of phonon modes most strongly associated to the motion of the Cu-O-Cu bond (3.3 and 6.8 THz) [30] and Ge bond (5 – 18 THz) are too large to account for the period of ≈ 1 ps (1 THz) that our data suggest. It is possible that the low energy mode at 0.9 THz discussed above may account for this behavior. To fully validate such a scenario will require further in-depth studies using next generation x-ray FEL sources.

Based on our current findings we postulate the following speculative picture of the dynamics following excitation, as outlined in Fig. 5. Our measurements are performed at a temperature of 20 K. Although this temperature is higher than $T_{SP}$, short range order fluctuations towards the spin-Peierls dimerization and concomitant singlet spin pairing are strong[32,33,53]. Within this environment, the intense ultraviolet pulse of 37.4 mJ/cm$^2$ excites the electronic sub-system and creates a charge transfer from the O to the Cu ions. This leads to a spatial redistribution of the electronic density in the CuO$_2$ chains of CuGeO$_3$ at 0 ps and acts as a trigger for launching a coherent oscillation with a period of < 2 ps, as observed in the spin response detected by RIXS via the ZRS excitation. We speculate here that this oscillation corresponds to the excitation of fluctuating magneto-elastic quasiparticles at very low energy. By modulating the bound magnon pairs via the spin-phonon coupling, this therefore results in a temporal modulation of the short-range AFM order, which is reflected in the ZRS intensity. Since the order is short-ranged, the coherent oscillation is strongly damped. At the same time, energy is efficiently transferred from the electronic into the lattice sub-system, raising the quasi-temperature of the lattice. As the quasi-temperature of the lattice increases, the magneto-elastic coupling gradually disappears. The rising lattice temperature has likely two consequences: (i) it contributes further to the damping of the coherent oscillation and (ii) the energy transfer into the magnetic sub-system becomes less efficient, because the suppression of the magneto-elastic quasiparticles closes a coupling channel between the two sub-systems.

Our time-resolved RIXS study uncovers intriguing physics in the non-equilibrium dynamics of a quasi-one-dimensional cuprate, CuGeO$_3$, allowing us to elaborate a possible scenario for their ultrafast evolution on the ps time scale. It calls for further experiments, with a more systematic approach regarding fluence and temperature dependences in particular. This will be made possible with next generation x-ray free electron facilities, allowing for higher statistics and for the acquisition of extensive data sets within a few days. In parallel, it would be important to reveal the ultrafast dynamics of the lattice degrees of freedom of CuGeO$_3$ directly. For instance, a future experiment using time-resolved x-ray diffraction to monitor the structural distortion related to the spin-Peierls phase would be highly beneficial.

In summary, by allowing access to the short-range magnetic correlations in CuGeO$_3$, trRIXS provides us with a powerful tool to probe the ultrafast dynamics of the local spin arrangement. Our current study outlines the complex interplay between the electronic, lattice and magnetic degrees of freedom in CuGeO$_3$. This establishes trRIXS as a technique capable of resolving the femtosecond dynamics of short-range magnetic correlations in low-dimensional and frustrated materials.

## Methods

**Sample preparation and RIXS characterization.** $CuGeO_3$ single crystals were cleaved at room pressure and temperature, producing mirror-like surfaces, and quickly transferred into the vacuum chamber (base pressure $10^{-9}$ mbar). The surface is oriented perpendicular to the [100] axis, so that the $CuO_4$ plaquettes are tilted 56° away from the surface. RIXS experiments were performed at the ADRESS beamline[54] of the Swiss Light Source, Paul Scherrer Institut, using the SAXES spectrometer[55]. A scattering angle of 90° was used and all the spectra were measured at the specular position, meaning that no light momentum is transferred to the system along the chain direction. The combined energy resolution was 60 meV at the oxygen K edge (~ 530 eV). Further details on the static RIXS measurements can be found in Section S3 of the Supplementary Materials.

**Time-resolved RIXS (trRIXS).** Measurements were carried out at the SXR beamline of the Linac Coherent Light Source operating at 120 Hz[56]. The system is excited using a 4.7 eV ultraviolet laser pulse of 50 fs duration generated by frequency addition in non-linear optical crystals. The energy of the 70-fs FEL x-ray pulses was tuned to the O *K*-edge (531 eV). The radiation scattered from the sample is collected by a compact spectrometer placed at a 90° scattering angle. The pump and probe pulses propagated collinearly and impinge with an angle of 45° on the sample surface. Both the laser pump pulse and the FEL probe pulse are horizontally polarized, i.e. the polarization vector lies in the scattering plane. In the experimental geometry, the $CuO_4$ plaquettes of copper-oxygen chains of $CuGeO_3$ lie at an angle of 56° with respect to the cleavage plane. The sample is kept at a temperature of 20 K during all the measurements. Further details on the trRIXS measurements can be found in Section S1 of the Supplementary Materials.

## References

1. Radu, I. *et al.* Transient ferromagnetic-like state mediating ultrafast reversal of antiferromagnetically coupled spins. *Nature* **472**, 205–208 (2011).

2. Vaterlaus, A., Beutler, T., Guarisco, D., Lutz, M. & Meier, F. Spin-lattice relaxation in ferromagnets studied by time-resolved spin-polarized photoemission. *Phys. Rev. B* **46**, 5280 (1992).

3. Stamm, C. *et al.* Femtosecond modification of electron localization and transfer of angular momentum in nickel. *Nat. Mater.* **6**, 740–743 (2007).

4. Wietstruk, M. *et al.* Hot-electron-driven enhancement of spin-lattice coupling in Gd and Tb 4f ferromagnets observed by femtosecond x-ray magnetic circular dichroism. *Phys. Rev. Lett.* **106**, 127401 (2011).

5. Maehrlein, S. F. *et al.* Dissecting spin-phonon equilibration in ferrimagnetic insulators by ultrafast lattice excitation. *Sci. Adv.* **4**, eaar5164 (2018).

6. Beaurepaire, E., Merle, J. C., Daunois, A. & Bigot, J. Y. Ultrafast spin dynamics in ferromagnetic nickel. *Phys. Rev. Lett.* **76**, 4250 (1996).

7. Carley, R. *et al.* Femtosecond laser excitation drives ferromagnetic gadolinium out of magnetic equilibrium. *Phys. Rev. Lett.* **109**, 057401 (2012).

8. Nicholson, C. W. *et al.* Ultrafast spin density wave transition in chromium governed by

thermalized electron gas. *Phys. Rev. Lett.* **117**, 136801 (2016).

9. Eich, S. *et al.* Band structure evolution during the ultrafast ferromagnetic-paramagnetic phase transition in cobalt. *Sci. Adv.* **3**, e1602094 (2017).

10. Tengdin, P. *et al.* Critical behavior within 20 fs drives the out-of-equilibrium laser-induced magnetic phase transition in nickel. *Sci. Adv.* **4**, eaap9744 (2018).

11. Mermin, N. D. & Wagner, H. Absence of ferromagnetism or antiferromagnetism in one- or two-dimensional isotropic Heisenberg models. *Phys. Rev. Lett.* **17**, 1133–1136 (1966).

12. Ament, L. J. P., Van Veenendaal, M., Devereaux, T. P., Hill, J. P. & Van Den Brink, J. Resonant inelastic x-ray scattering studies of elementary excitations. *Rev. Mod. Phys.* **83**, 705–767 (2011).

13. Schlappa, J. *et al.* Spin-orbital separation in the quasi-one-dimensional Mott insulator Sr2CuO3. *Nature* **485**, 82–85 (2012).

14. Le Tacon, M. *et al.* Intense paramagnon excitations in a large family of high-temperature superconductors. *Nat. Phys.* **7**, 725–730 (2011).

15. Wang, Y., Chen, Y., Jia, C., Moritz, B. & Devereaux, T. P. Time-resolved resonant inelastic x-ray scattering in a pumped Mott insulator. *Phys. Rev. B* **101**, 165126 (2020).

16. Dean, M. P. M. *et al.* Ultrafast energy- and momentum-resolved dynamics of magnetic correlations in the photo-doped Mott insulator Sr2IrO4. *Nat. Mater.* **15**, 601-605 (2016).

17. Mitrano, M. *et al.* Ultrafast time-resolved x-ray scattering reveals diffusive charge order dynamics in La2-xBaxCuO4. *Sci. Adv.* **5**, eaax3346 (2019).

18. Wernet, P. *et al.* Orbital-specific mapping of the ligand exchange dynamics of Fe(CO)5 in solution. *Nature* **520**, 78–81 (2015).

19. Parchenko, S. *et al.* Orbital dynamics during an ultrafast insulator to metal transition. *Phys. Rev. Res.* **2**, 23110 (2020).

20. Mitrano, M. & Wang, Y. Probing light-driven quantum materials with ultrafast resonant inelastic X-ray scattering. *Commun. Phys.* **3**, 184 (2020).

21. Völlenkle, H., Wittmann, A. & Nowotny, H. Zur Kristallstruktur von CuGeO3. *Monatshefte für Chemie* **98**, 1352–1357 (1967).

22. Braden, M. *et al.* Structural analysis of CuGeO3: Relation between nuclear structure and magnetic interaction. *Phys. Rev. B* **54**, 1105–1116 (1996).

23. Hase, M., Terasaki, I. & Uchinokura, K. Observation of the spin-Peierls transition in linear Cu2+ (spin-1/2) chains in an inorganic compound CuGeO3. *Phys. Rev. Lett.* **70**, 3651–3654 (1993).

24. Hase, M. *et al.* Magnetic phase diagram of the spin-Peierls cuprate CuGeO3. *Phys. Rev. B* **48**, 9616–9619 (1993).

25. Hirota, K. *et al.* Dimerization of CuGeO3 in the spin-Peierls state. *Phys. Rev. Lett.* **73**, 736–739 (1994).

26. Castilla, G., Chakravarty, S. & Emery, V. J. Quantum magnetism of CuGeO3. *Phys. Rev. Lett.* **75**, 1823–1826 (1995).

27. Braden, M., Reichardt, W., Hennion, B., Dhalenne, G. & Revcolevschi, A. Lattice dynamics of CuGeO3: Inelastic neutron scattering and model calculations. *Phys. Rev. B* **66**, 214417 (2002).

28. Seidel, A., Marianetti, C. A., Chou, F. C., Ceder, G. & Lee, P. A. S = 1/2 chains and spin-Peierls transition in TiOCl. *Phys. Rev. B* **67**, 20405 (2003).

29. Abel, E. T. *et al.* X-ray scattering study of the spin-Peierls transition and soft phonon behavior in TiOCl. *Phys. Rev. B* **76**, 214304 (2007).

30. Braden, M., Hennion, B., Reichardt, W., Dhalenne, G. & Revcolevschi, A. Spin-phonon coupling in CuGeO3. *Phys. Rev. Lett.* **80**, 3634–3637 (1998).

31. Cowley, R. A. Structural phase transitions I. Landau theory. *Adv. Phys.* **29**, 1–110 (1980).

32. Chen, C. H. & Cheong, S.-W. Lattice fluctuations well above the spin-Peierls transition in the linear-chain system CuGeO3. *Phys. Rev. B* **51**, 6777–6779 (1995).

33. Schoeffel, J. P., Pouget, J. P., Dhalenne, G. & Revcolevschi, A. Spin-Peierls lattice fluctuations of pure and Si- and Zn-substituted CuGeO3. *Phys. Rev. B* **53**, 14971–14979 (1996).

34. Hirota, K. *et al.* Characterization of the structural and magnetic fluctuations near the spin-Peierls transition in CuGeO3. *Phys. Rev. B* **52**, 15412–15419 (1995).

35. Monney, C. *et al.* Determining the short-range spin correlations in the spin-chain Li2CuO2 and CuGeO3 compounds using resonant inelastic x-ray scattering. *Phys. Rev. Lett.* **110**, 087403 (2013).

36. Duda, L.-C. *et al.* Bandlike and excitonic states of oxygen in CuGeO3: Observation using polarized resonant soft-x-ray emission spectroscopy. *Phys. Rev. B* **61**, 4186–4189 (2000).

37. Monney, C. *et al.* Probing inter- and intrachain Zhang-Rice excitons in Li2CuO2 and determining their binding energy. *Phys. Rev. B* **94**, 165118 (2016).

38. Okada, K. & Kotani, A. Zhang-Rice singlet-state formation by oxygen 1s resonant x-ray emission in edge-sharing copper-oxide systems. *Phys. Rev. B* **63**, 45103 (2001).

39. Zagoulaev, S. & Tupitsyn, I. I. Electronic structure and magnetic properties of the spin-Peierls compound CuGeO3. *Phys. Rev. B* **55**, 13528–13541 (1997).

40. Pagliara, S., Parmigiani, F., Galinetto, P., Revcolevschi, A. & Samoggia, G. Role of the Zhang-Rice-like exciton in optical absorption spectra of CuGeO3 and CuGe1-xSixO3 single crystals. *Phys. Rev. B* **66**, 24518 (2002).

41. Giannetti, C. *et al.* Disentangling thermal and nonthermal excited states in a charge-transfer insulator by time- and frequency-resolved pump-probe spectroscopy. *Phys. Rev. B* **80**, 235129 (2009).

42. Marciniak, A. *et al.* Vibrational coherent control of localized d–d electronic excitation. *Nat. Phys.* **17**, 368-373 (2021).

43. Masayuki, U. *et al.* Raman scattering of CuGeO3. *J. Phys. Soc. Jpn.* **63**, 4060–4064 (1994).

44. Kuroe, H. *et al.* Raman-scattering study of CuGeO3 in the spin-Peierls phase. *Phys. Rev.*

*B* **50**, 16468–16474 (1994).

45. van Loosdrecht, P. H. M., Boucher, J. P., Martinez, G., Dhalenne, G. & Revcolevschi, A. Inelastic light scattering from magnetic fluctuations in CuGeO3. *Phys. Rev. Lett.* **76**, 311–314 (1996).

46. Muthukumar, V. N. *et al.* J1-J2 model revisited: Phenomenology of CuGeO3. *Phys. Rev. B* **55**, 5944–5952 (1997).

47. Loa, I., Gronemeyer, S., Thomsen, C. & Kremer, R. K. Spin gap and spin-phonon interaction in CuGeO3. *Solid State Commun.* **99**, 231–235 (1996).

48. Kuroe, H. *et al.* Raman-scattering study of CuGeO3. *Physica B* **219–220**, 104–106 (1996).

49. Ogita, N. *et al.* Raman scattering of CuGeO3. *Physica B* **219–220**, 107–109 (1996).

50. Els, G. *et al.* Observation of three-magnon light scattering in CuGeO3. *Phys. Rev. Lett.* **79**, 5138–5141 (1997).

51. Uhrig, G. S. & Schulz, H. J. Magnetic excitation spectrum of dimerized antiferromagnetic chains. *Phys. Rev. B* **54**, R9624--R9627 (1996).

52. Geertsma, W. & Khomskii, D. Influence of side groups on 90° superexchange: A modification of the Goodenough-Kanamori-Anderson rules. *Phys. Rev. B* **54**, 3011–3014 (1996).

53. Holicki, M., Fehske, H. & Werner, R. Magnetoelastic excitations in spin-Peierls systems. *Phys. Rev. B* **63**, 174417 (2001).

54. Strocov, V. N. *et al.* High-resolution soft x-ray beamline ADRESS at the swiss light source for resonant inelastic x-ray scattering and angle-resolved photoelectron spectroscopies. *J. Synchrotron Radiat.* **17**, 631–643 (2010).

55. Ghiringhelli, G. *et al.* SAXES, a high resolution spectrometer for resonant x-ray emission in the 400-1600 eV energy range. *Rev. Sci. Instrum.* **77**, 113108 (2006).

56. Schlotter, W. F. *et al.* The soft x-ray instrument for materials studies at the linac coherent light source x-ray free-electron laser. *Rev. Sci. Instrum.* **83**, 43107 (2012).

**Acknowledgments**

The static synchrotron experiments have been performed at the ADRESS beamline of the Swiss Light Source at the Paul Scherrer Institut (PSI). Work at PSI was funded by the Swiss National Science Foundation through the Sinergia project "Mott Physics Beyond the Heisenberg (MPBH) model" (SNSF Research grant numbers CRSII2_141962 and CRSII2_160765). This project was also supported from the Swiss National Science Foundation (SNSF) Grant No. P00P2_170597. S. J. acknowledges support from the National Science Foundation under Grant No. DMR-1842056. Use of the Linac Coherent Light Source (LCLS), SLAC National Accelerator Laboratory, is supported by the U.S. Department of Energy, Office of Science, Office of Basic Energy Sciences under Contract No. DE-AC02-76SF00515.

**Author contributions**

G.D., C.M. and T.S. conceived the project and coordinated the project phases. E.P., C.W.N, Y.T., M.R. and T.S. performed the static RIXS experiments at SLS/PSI with the assistance of V.N.S.. R.S.-M. and A.R. synthesized the samples. E.P., C.W.N, Y.T., G.C., S.Z., M.F.L., W.S., G.D., C.M. and T.S. carried out the trRIXS experiments at LCLS/SLAC. E.P., C.W.N, S.Z. and C.M. analyzed the data in discussion with T.S.. A.K. and S.J. performed the theoretical simulations. E.P., C.W.N, C.M. and T.S. wrote the manuscript with input from all authors. E.P. and C.W.N. contributed equally to this work.

**Competing financial interests**

The authors declare no competing interests.

**Data and materials availability**

All the data used in the present manuscript are available on request.

**Figure Captions**

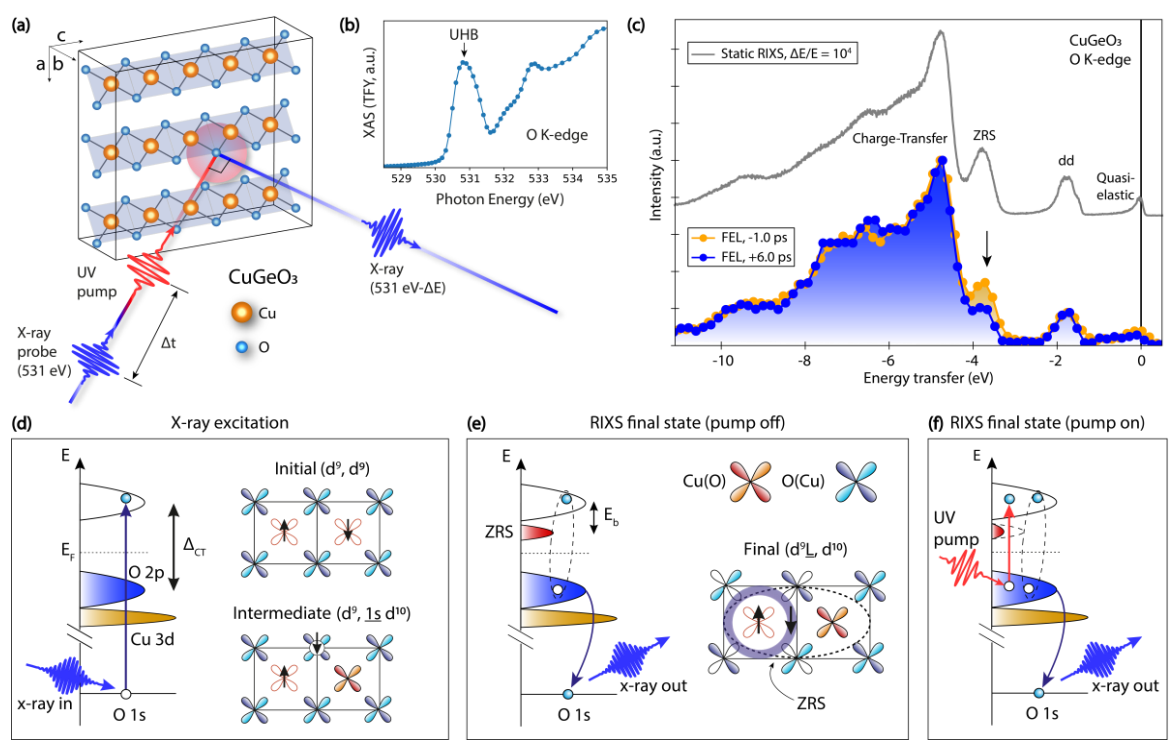

**Fig. 1. Experimental overview and schematic of ZRS formation.** (a) Sketch of the experimental scheme at the SXR beamline of the LCLS x-ray free-electron laser (FEL) facility. The laser pump pulse (4.7 eV, 50 fs) is followed by a collinear x-ray probe pulse (531 eV, 70 fs). The radiation re-emitted by the sample is then analyzed using a grating-based spectrometer. (b) Static O K-edge XAS spectrum of $CuGeO_3$, taken at 20 K using a synchrotron light source. The data is acquired in the same experimental geometry as the FEL experiment. (c) A comparison between the O *K*-edge RIXS spectrum obtained at a synchrotron radiation facility

with a resolving power $10^4$ (shown with a vertical offset) and those obtained with FEL radiation for a negative delay (yellow markers) and a positive delay (blue markers). Fluence = 37.4 mJ/cm². The black arrow marks the position of the Zhang-Rice singlet (ZRS) excitation. (d)-(f) Schematics of the ZRS formation in the O *K*-edge RIXS process described in the main text. In the schematics, $E_F$ is the Fermi level, $\Delta_{CT}$ is the charge-transfer energy, and $E_b$ is the ZRS binding energy.

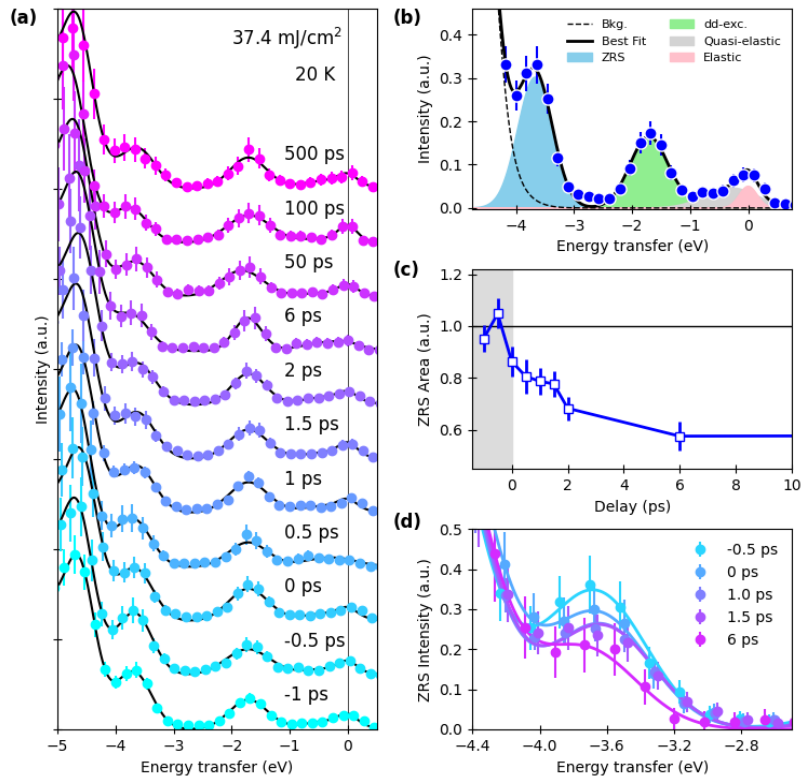

**Fig. 2. Time-resolved RIXS spectra and ZRS dynamics.** (a) O *K*-edge RIXS spectra (markers) collected with a pump fluence of 37.4 mJ/cm². Spectra taken at different pump-probe delays are shown with a vertical offset. The black line represents the best fit curve. The error bars are obtained by analyzing the fluctuation between different acquisitions taken under the same conditions. (b) Example of the multi-component fit procedure used to extract the intensity of the ZRS excitation. The different components (denoted in the legend) are modeled with Gaussian functions. (c) Evolution of the integrated intensity of the ZRS excitation as a function of the pump-probe delay. The error bars represent the confidence interval obtained from the fit. (d) O *K*-edge RIXS spectrum around the ZRS excitation for selected time delays as indicated in the legend.

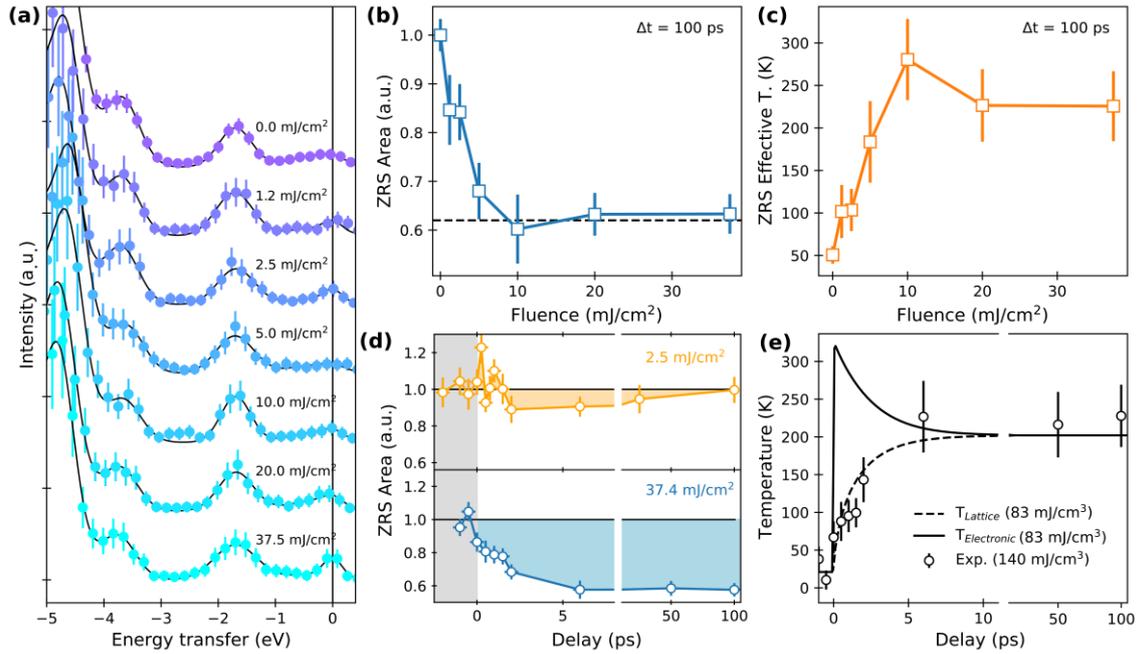

**Fig. 3. Fluence dependent dynamics and effective temperature.** (a) O $K$-edge RIXS spectra (markers) presented as a function of the pump fluence at $\Delta t = 100$ ps. Spectra are shown with a vertical offset for clarity. The black solid line represents the best fit curve. (b) Fluence dependence of the integrated intensity of the ZRS excitation, exhibiting saturation for $F > 5$ mJ/cm$^2$ marked by the dashed line. The error bars represent the confidence interval from the least-square fit. (c) Fluence dependence of the effective ZRS temperature, extrapolated using the static temperature dependence of the ZRS (see section S3 of the Supplemental Materials). The saturation of the ZRS signal occurs below the room temperature value. (d) Time trace of the integrated intensity of the ZRS for two different laser fluences. Data are normalized to the average value at negative time delays. Data at different fluence is shown with separated panels for clarity. (e) Two-temperature model describing the energy transfer between the electronic and lattice subsystems (see Section S3 in the Supplementary Materials). The solid line shows the electronic temperature, the dashed line the lattice temperature, while the white markers are experimental points. The absorbed heat content estimated for the experiment and the one used in the calculation are indicated in brackets in the legend.

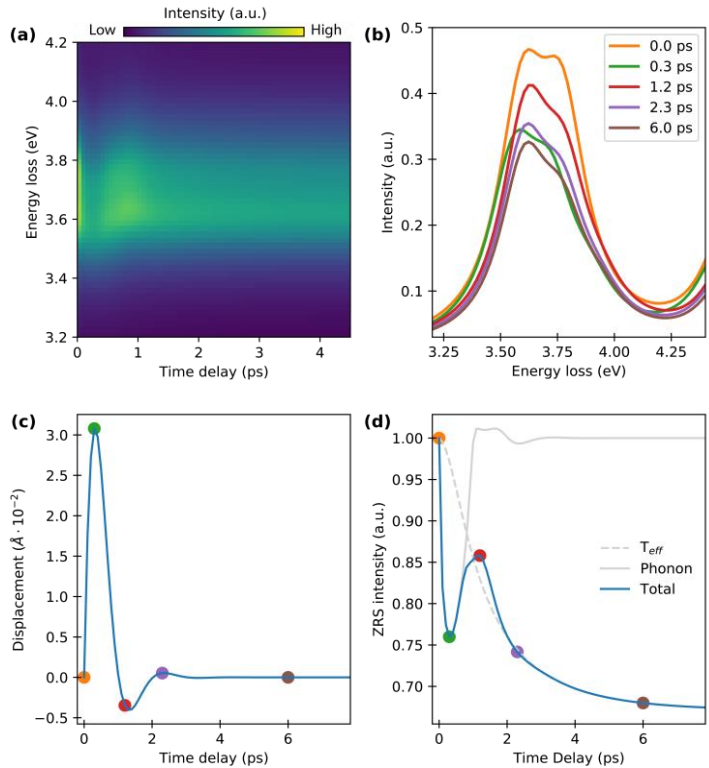

**Fig. 4. Calculated trRIXS spectra in the region of the ZRS.** (a) O $K$-edge RIXS in the vicinity of the ZRS excitation as a function of the time delay, calculated using the $Cu_3O_8$ cluster model. (b) Calculated O $K$-edge RIXS spectra at selected time delays. (c) Time evolution of the atomic displacement used in the calculation, including a phenomenological exponential damping. (d) Computed time evolution of the ZRS intensity. The solid and dashed gray lines represent the effect of the lattice vibration and of the effective temperature on the computed ZRS intensity, respectively. The colored dots in the time traces of panels (c-d) indicate the time delays chosen for the calculated RIXS spectra shown in panel (b), with the same color code. Full details of the calculations in the Supplementary Materials, Section S4.

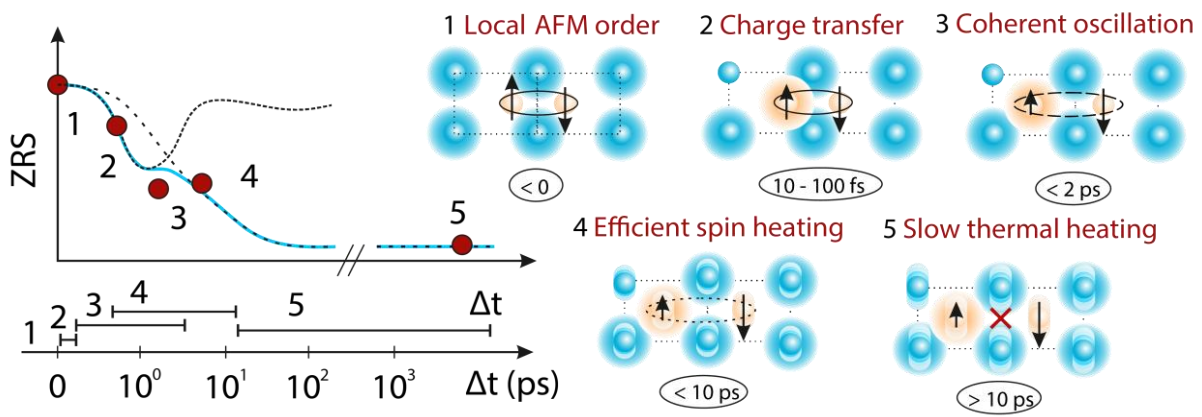

**Fig. 5. Schematics of microscopic processes following photoexcitation.** (Left) Approximate time scales for the different processes described (right). The pump pulse causes a charge transfer between O and Cu, which launches a short-range oscillation of a magneto-elastic mode,

resulting in a modulation of the AFM coupling strength and correspondingly the ZRS intensity. Due to the heat transfer to the lattice, this mode is gradually removed and therefore the lattice and magnetic temperatures separate.

# Supplementary materials for

# Probing the interplay between lattice dynamics and local magnetic correlations in CuGeO$_3$ with femtosecond RIXS

E. Paris[1,†,*], C. W. Nicholson[2,†], S. Johnston[3], Y. Tseng[1], M. Rumo[2], G. Coslovich[4], S. Zohar[4], M.F. Lin[4], V.N. Strocov[1], R. Saint-Martin[5], A. Revcolevschi[5], A. Kemper[6], W. Schlotter[4], G. L. Dakovski[4], C. Monney[2], and T. Schmitt[1,*]

[1]*Photon Science Division, Paul Scherrer Institut, CH-5232 Villigen PSI, Switzerland*

[2]*Départment de Physique and Fribourg Centre for Nanomaterials, University of Fribourg, CH-1700 Fribourg, Switzerland*

[3]*Department of Physics and Astronomy, University of Tennessee, Knoxville, TN 37996 USA*

[4]*Linac Coherent Light Source, SLAC National Accelerator Laboratory, Menlo Park, CA 94025, USA*

[5]*Laboratoire de Physico-Chimie de l'Etat Solide, ICMMO, Université Paris-Saclay, 91405 Orsay Cedex, France*

[6]*Department of Physics, North Carolina State University, Raleigh, NC 27695, USA*

†Equally contributing authors

*Corresponding authors. Email: eugenio.paris@psi.ch (E.P.); thorsten.schmitt@psi.ch (T.S.)

## Supplementary text

### Section S1: Experimental setup and data acquisition

The optical pump, soft x-ray probe experiment was performed at the SXR beamline of the Linac Coherent Light Source (LCLS) free-electron laser (FEL) facility[1]. The energy of the π-polarized x-ray pulses, with pulse duration of 70 fs and repetition rate of 120 Hz, was tuned to the O $K$-edge (~ 531 eV). The energy bandwidth was reduced using a plane-grating monochromator down to $\Delta E$ = 100 meV. The 4.7 eV (266 nm) pump pulses, with 50 fs duration, were generated by frequency tripling the fundamental of a Ti:Sapphire laser operated at 120 Hz and focused to a 100 x 500 µm$^2$ elliptical spot (FWHM) at the sample. The x-ray beam was focused using Kirkpatrick-Baez optics to reach a spot at the sample of 30 µm along the vertical direction. The horizontal size of the x-ray beam was adjusted to be contained in the laser spot size. The penetration depth of the soft x-rays is ~ 100 nm (1/e attenuation depth) across the O $K$-edge[2], as opposed to the ~ 2.1 µm penetration depth (1/e attenuation depth) of the pump pulse [3,4], ensures that the total volume probed by the x-rays is homogeneously pumped. The beams were spatially overlapped onto a frosted Ce:YAG crystal and synchronized by monitoring the transmission changes induced by the x-rays on a transparent YAG crystal. The temporal jitter between pump and probe was measured by means of a timing-tool on a shot-by-shot basis. However, the limitations in terms of statistics did not allow reducing the time resolution better than ~ 400 fs by event sorting. The pump-probe time delay was selected using a mechanical translation stage.

The pump and probe beams propagate collinearly to the sample with 45° incidence angle from the sample surface and the scattered photons are detected using a soft x-ray spectrometer placed at a scattering angle of 90°. Both the laser pump pulse and the FEL probe pulse are horizontally polarized such that the polarization vector lies in the scattering plane. We estimate the combined energy resolution to be 365 meV FWHM, obtained with the spectrometer grating set to the 1st diffraction order. The spectrometer disperses the scattered x-rays vertically through a variable line-spacing grating, allowing it to collect a larger solid angle in the horizontal plane. This allows using an extended horizontal line focus that reduces the sample damage by the x-rays without increasing the time required to collect the signal. The spectrometer was equipped with an ANDOR Newton DO940P CCD camera operated at 120 Hz readout rate in 1D binning mode along the non-dispersive direction. The single-crystal $CuGeO_3$ sample, synthesized as discussed elsewhere[5], was placed on a manipulator equipped with a continuous-flow Helium cryostat to maintain the sample at a constant temperature of 20 K during all measurements.

Most of the data have been collected using an incident laser fluence of 37.4 mJ/cm². The use of such a large fluence is due to the mismatch between the penetration depth of soft x-rays and UV light. Indeed, assuming an exponential absorption profile in the sample, within the x-ray probing depth of 100 nm only ~5% of the total pump fluence is absorbed. This results in a total absorbed fluence of ~140 J/cm³ in the volume probed by the x-rays. Assuming that every pump photon is absorbed by the electronic system, this leads to the excitation of ~0.023 electrons per unit cell.

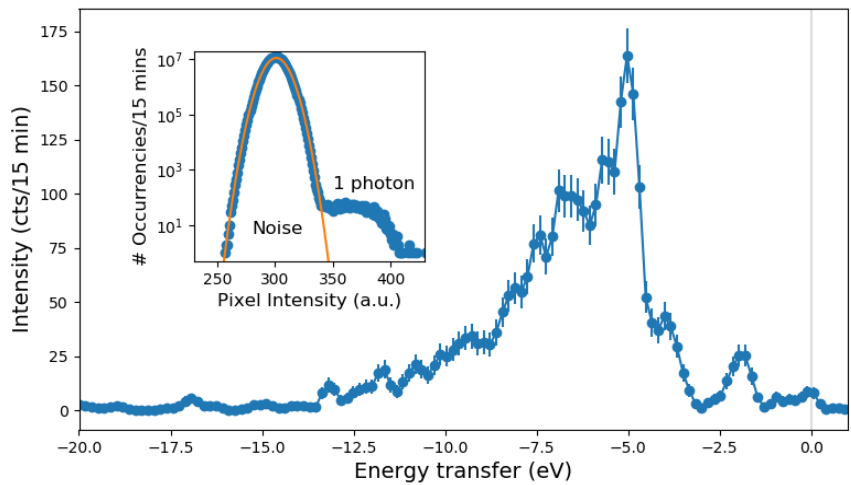

**Figure S 1: Typical acquisition statistics with the RIXS setup** described in the text. (Main) O K-edge RIXS spectrum obtained in a 15 minutes accumulation as obtained using the single-photon counting algorithm. The error bars are given as the square root of the number of counts. (Inset) Histogram of the pixel intensity of the CCD camera in a 15-minutes accumulation. The data are shown as blue dots. The orange solid line is a Gaussian fit of the noise peak.

The inset of Fig. S1 shows a histogram of the intensity on the CCD camera in 15 minutes acquisition of the RIXS spectrum. The RIXS signal is partially separated from the dominant noise contribution, appearing as a large Gaussian peak. A single photon counting scheme, based on a 1D centroiding algorithm is then applied to the raw data, allowing a better separation of the 1-photon signal from the noise. Due to the limitations in photon statistics, we set the number of points in the centroid algorithm to be the same as the number of physical pixels in the CCD camera.

# Section S2: Fitting of the RIXS spectra

In what follows, we present the fitting procedure used to extract the evolution of the Zhang-Rice singlet as a function of the time delay and the pump fluence. In our experiment, we typically acquire several runs of 15 minutes each at a given time delay. We then average them to obtain the final RIXS spectrum. In the following, the error bars associated to every experimental point are obtained by evaluating the variability between different 15-minutes accumulations under the same experimental conditions. The data were then normalized with respect to the total area under the RIXS spectrum to minimize possible artefacts resulting from intensity fluctuations due to experimental instabilities during the acquisition. Therefore, we assume that the total integrated emission from the sample presents only negligible changes as a function of the pump-probe delay. After normalization, we fit the RIXS spectra as shown in Figs. S2, S3, and S4. To reproduce the elastic line, we use a Gaussian function with the width fixed at the instrumental resolution. The asymmetry due to the quasi-elastic scattering is modelled using an additional Gaussian function with fixed position and width. Gaussian functions are used to describe the d-d excitations and the Zhang-Rice singlet (ZRS) excitation. The charge-transfer (CT) contribution, appearing as a complex structure extending from 4 eV to 15 eV energy loss, is modelled as a combination of six Gaussian functions with fixed widths. The relative energy position of each component of the CT, as well as its relative amplitude is fixed. Therefore, only a global shift and a global intensity variation is allowed for the CT, which accounts for fluctuations in the background while preserving the overall shape of the CT. Due to the low visibility of the elastic line, a fine calibration of the energy scale has been performed using the position of the d-d excitations as a reference, under the assumption that the energy associated to the d-d excitations does not change due to the laser pump within our energy resolution. To ensure the fit stability, we choose to fix the width of the ZRS peak to its equilibrium value. The RIXS data, along with the best-fit curves, are shown at different time delay for $F = 37.4$ mJ/cm$^2$ (Fig. S2) and for $F = 2.5$ mJ/cm$^2$ (Fig. S3), and as a function of the fluence at 100 ps time delay (Fig. S4).

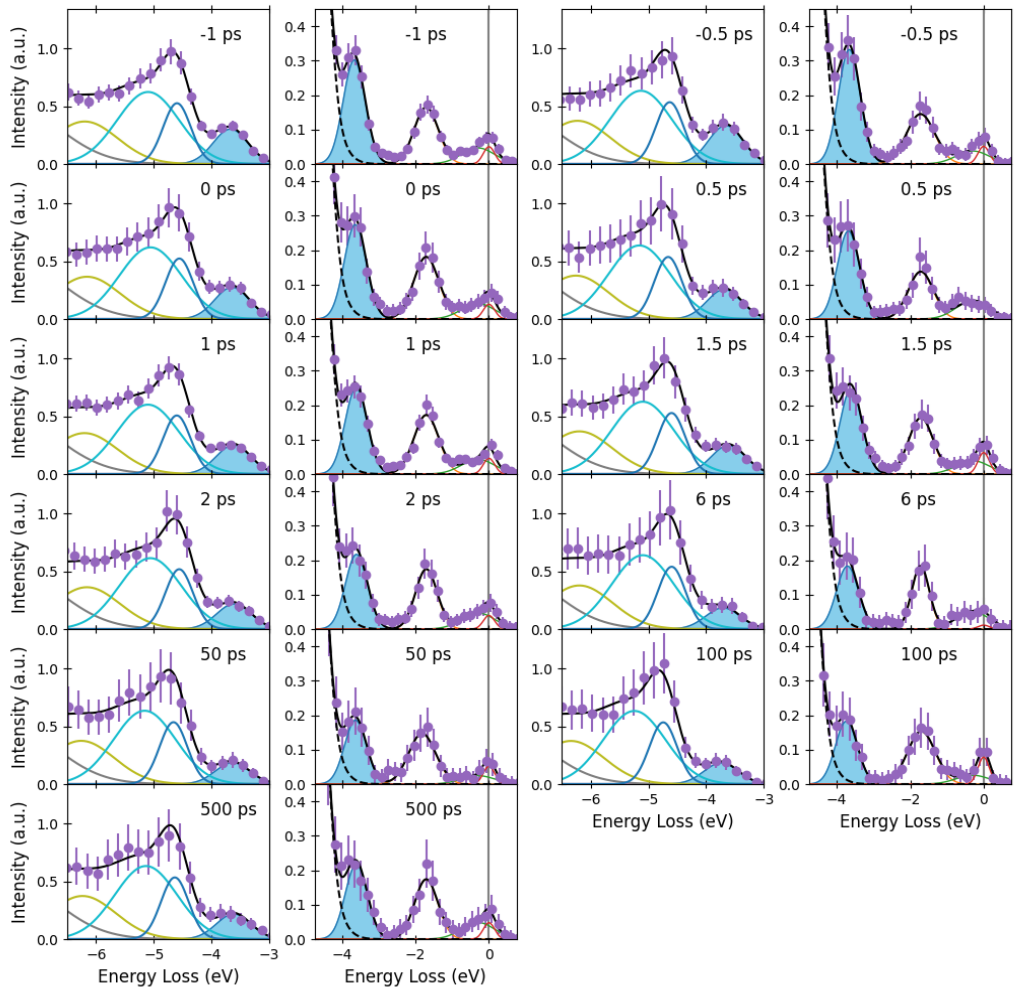

**Figure S 2: Fit of the O K-edge RIXS data at different time delays for a pump fluence of 37.4 mJ/cm$^2$.** The purple dots are the RIXS data while the best fit is presented as a black solid line. The time delay of each spectrum is specified in the panel. For each time delay, two different energy ranges are shown. In the left panels, the ZRS peak (blue filled) is shown along with the 4 lower-energy components of the CT background. In the right panels, we present the low-energy part of the RIXS spectrum. The single components are the elastic line (red), the quasi-elastic (green), the dd-excitations, the Zhang-Rice singlet (blue filled), and the overall contribution of the charge-transfer background (black dashed).

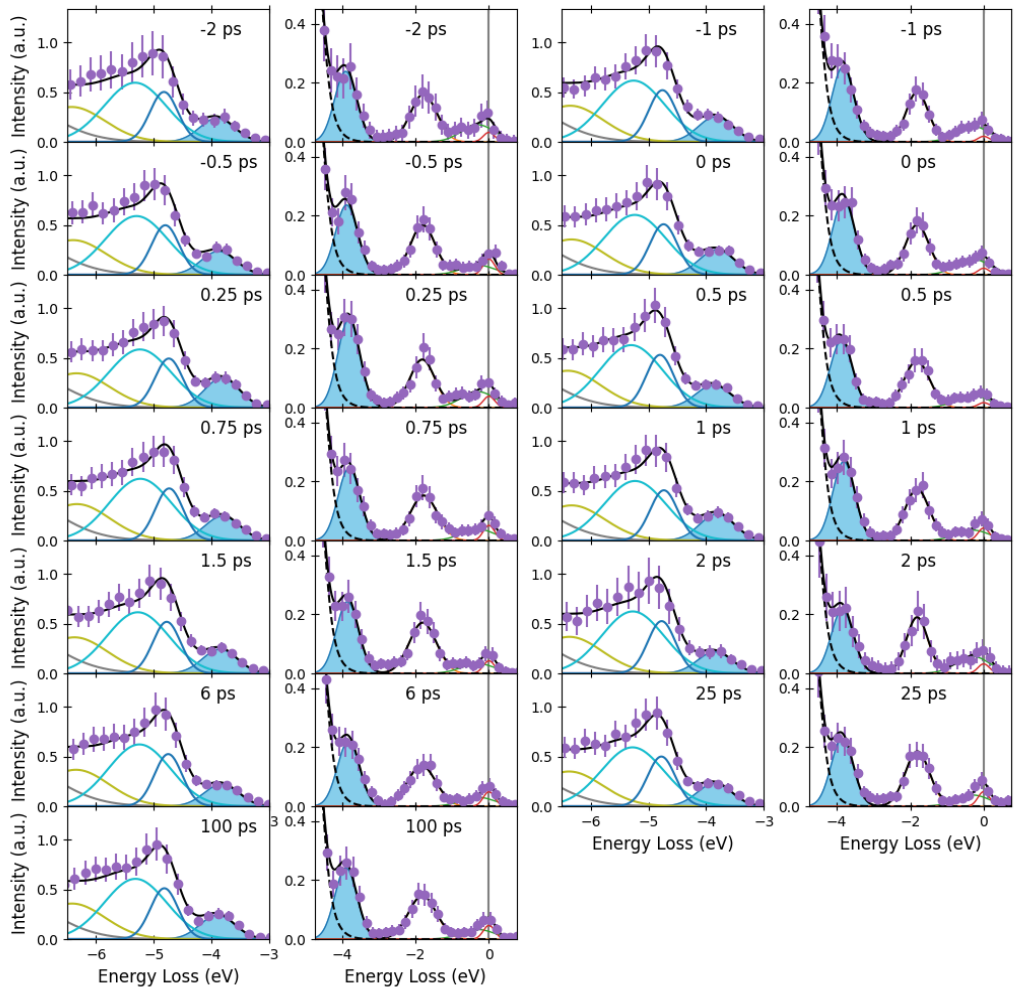

**Figure S 3: Fit of the O K-edge RIXS data at different time delays for a pump fluence of 2.5 mJ/cm$^2$**. The purple dots are the RIXS data while the best fit is presented as a black solid line. The time delay of each spectrum is specified in the panel. For each time delay, two different energy ranges are shown. In the left panels, the ZRS peak (blue filled) is shown along with the 4 lower-energy components of the CT background. In the right panels, we present the low-energy part of the RIXS spectrum. The single components are the elastic line (red), the quasi-elastic (green), the dd-excitations, the Zhang-Rice singlet (blue filled), and the overall contribution of the charge-transfer background (black dashed).

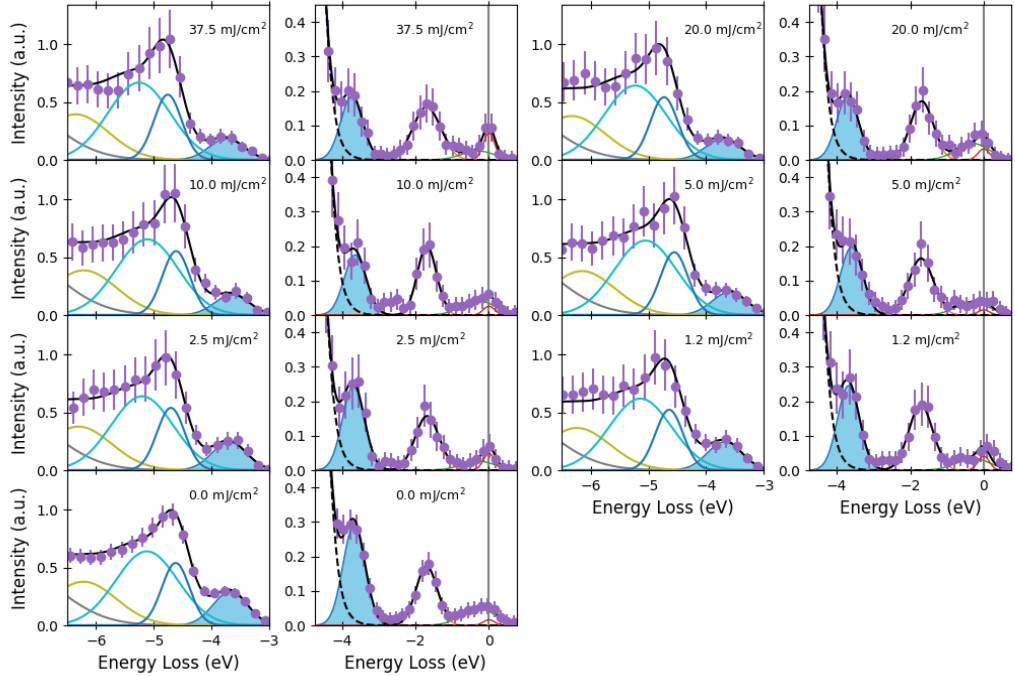

**Figure S 4: Fit of the O K-edge RIXS data take at 100 ps time delay and different pump fluences.** The purple dots are the RIXS data while the best fit is presented as a black solid line. The pump fluence of each spectrum is specified in the panel. For each fluience, two different energy ranges are shown. In the left panels, the ZRS peak (blue filled) is shown along with the 4 lower-energy components of the CT background. In the right panels, we present the low-energy part of the RIXS spectrum. The single components are the elastic line (red), the quasi-elastic (green), the dd-excitations, the Zhang-Rice singlet (blue filled), and the overall contribution of the charge-transfer background (black dashed).

## Section S2.1: Sampling of the time delays

The timing tool available at the LCLS allows us to obtain a measurement of the actual time delay for each single FEL shot. As the low throughput of the present experiment does not allow obtaining a full RIXS spectrum at each shot, we accumulate $\geq 10^5$ shots for each given time delay and laser pump fluence and choose a certain bin size in the time delay within which perform the summation. This choice affects the overall time resolution of the measurement. The data presented in the main manuscript and in the section above is obtained using a 400-fs grid that we judge appropriate to achieve significant statistics. In Fig. S5, we show the evolution of the ZRS intensity as a function of the time delay obtained by averaging the data on the 400-fs grid (blue symbols) and the data points obtained with a 200-fs grid (red symbols). The latter has twice as much time resolution, but nearly half the statistics. As compared to the 400-fs binning, the 200 fs binning shows a deeper suppression of the intensity at short time delays and might suggest that the actual oscillation frequency is higher than that obtained with the 400-fs grid data. However, the large fluctuations of the result, even in the region of negative time delays, suggests that the data points obtained with such sampling are limited by the low statistics.

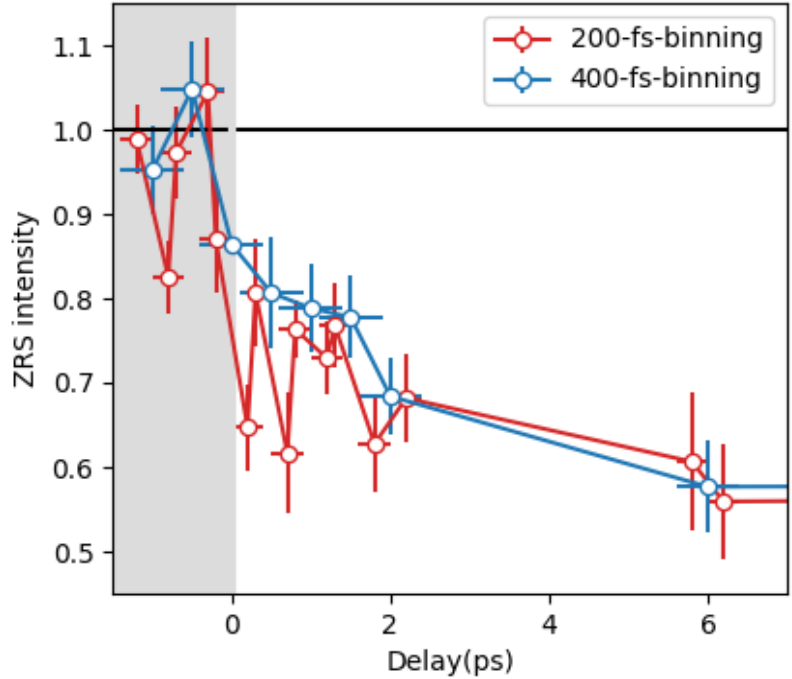

**Figure S 5: Evolution of the intensity of the ZRS excitation as a function of the pump-probe time delay at a fixed fluence of 37.4 mJ/cm$^2$.** The orange symbols represent the data collected with highest statistics. The blue symbols represent data collected with smaller time binning, but lower statistics.

### Section S2.2 Alternative fitting model

In section S2, we have presented the fitting strategy to extract the evolution of the ZRS excitation as a function of the pump-probe time delay and fluence. In such fitting approach (Model #1), we have kept the width of the ZRS peak fixed to the equilibrium value in order to avoid fit instabilities. In this section, we report the fit results obtained by letting the width of the ZRS free to adjust to find the best agreement with the experimental data (Model #2). The best fit curves for the two models are compared to the experimental data in Fig. S6 (a-b) while the result of the two fitting approaches is compared in Fig. S6 (c-d). The two fit models return nearly the same optimal values except for $\Delta t = 1$ ps and 1.5 ps, in which the width of the ZRS increases in Model #2 and so does the total integrated intensity. For instance, at $\Delta t = 1.5$ ps, the FWHM of the ZRS obtained with Model #2 is more than 300 meV larger than the one obtained with Model #1. As a result, the damped oscillation observed is strongly amplified. This result confirms the presence of a non-thermal state below 2 ps, however, further experiments with a finer energy sampling and better statistics are needed to characterize such non-thermal state in greater detail.

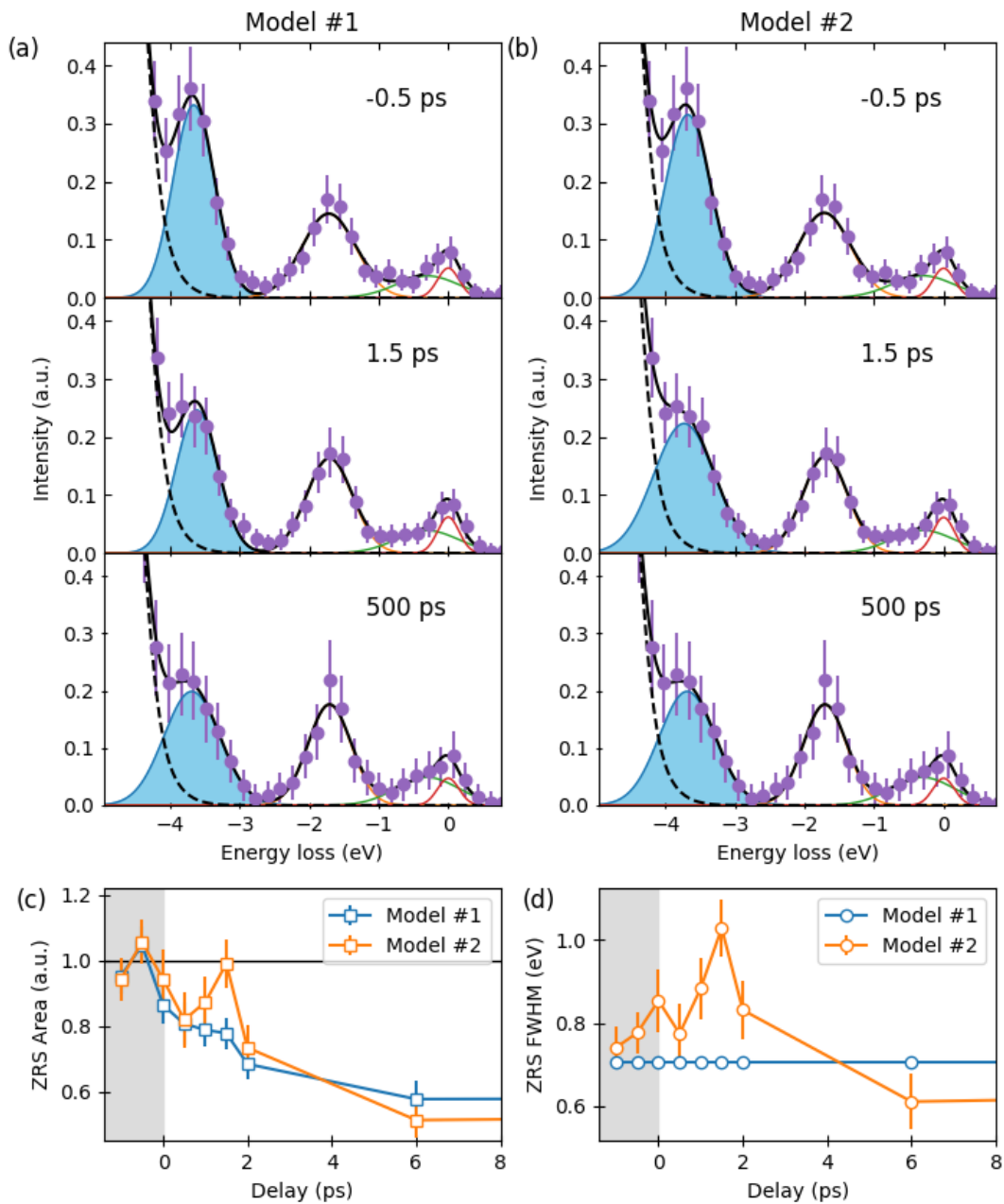

**Figure S 6: Comparison of the fit results obtained with two different models.** (a) Fit of the O K-edge RIXS data using Model #1 for three different time delays as indicated in the caption, with a pump fluence of 37.4 mJ/cm² (same data as shown in Fig. S2). (b) Fit results using Model #2. (c-d) best fit results for the area and FWHM of the ZRS excitation using the two different models.

## Section S3: Synchrotron RIXS measurements and two-temperature model analysis

The static O K-edge Resonant Inelastic X-ray Scattering (RIXS) measurements were carried out at the ADRESS beamline of the Swiss Light Source at the Paul Scherrer Institut, Switzerland. The x-ray beam was monochromatized using a plane grating and focused down to a spot size of ≤ 4x55 μm² at the sample position using an elliptical refocusing mirror. The incoming radiation was linearly polarized in the horizontal direction, i.e. with the polarization in the scattering plane. The scattering angle was set to 2θ

= 90° while all measurements were performed at the specular condition, i.e. with the incoming beam forming an angle of 45° with respect to the sample surface. The combined energy resolution was 55 meV FWHM, determined by collecting the elastic scattering from a carbon tape reference. The sample temperature was varied using a sample manipulator equipped with a continuous flow Helium cryostat.

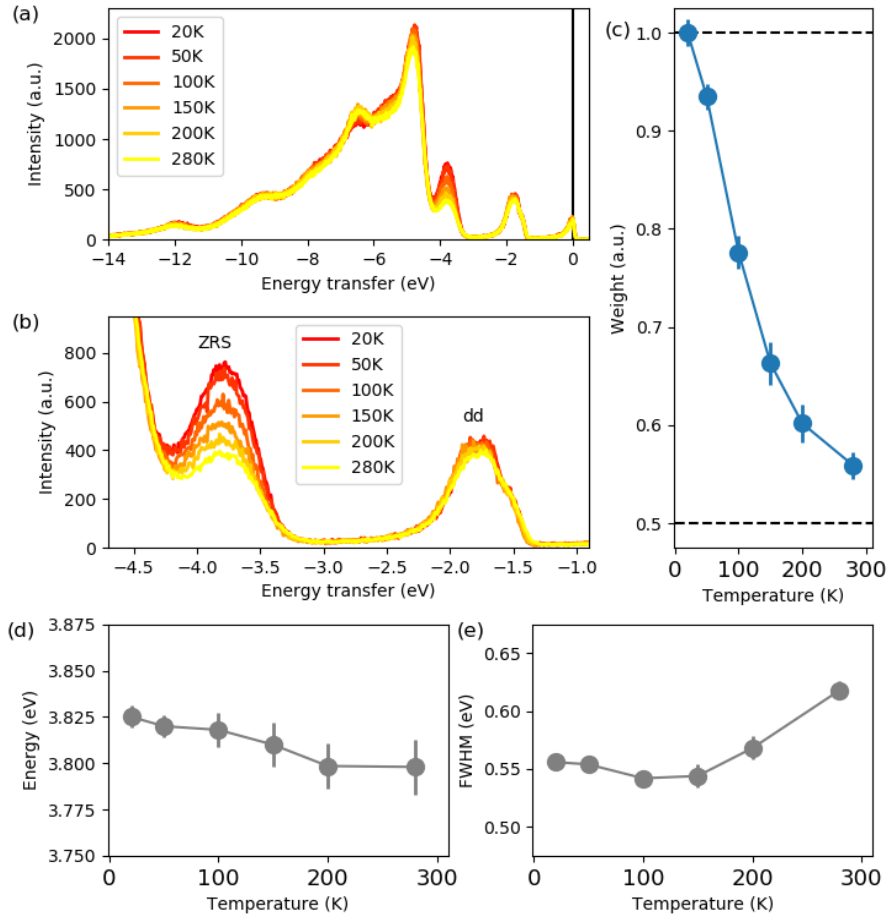

**Figure S 7: Temperature-dependent RIXS on CuGeO$_3$.** (a-b) Synchrotron O K-edge RIXS spectra measured at a synchrotron source at different temperatures. (c) Spectral weight, (d) position, and (e) FWHM of the Zhang-Rice singlet excitation as a function of temperature.

RIXS spectra of CuGeO$_3$ were obtained as a function of temperature between 20 K and 300 K, as show in Fig. S7 (a-b). This allowed the intensity evolution of the ZRS as a function of temperature to be systematically tracked (Fig. S7 (b-c)) under the same scattering geometry as utilized in our time-resolved measurements. Previously published results were obtained using a different geometry[6]. Due to the higher energy resolution available at the static experiment, the ZRS is well separated from the CT peak and its intensity vs temperature was extracted using a fitting procedure as described in the main text.

As outlined in the main text, a comparison is made between the static reduction of ZRS intensity caused by temperature and the temporal changes of the ZRS caused by the pump pulse in the time-resolved experiment. In this way, we obtain an estimate of the transient magnetic quasi-temperature. The transient ZRS intensity trace (normalized to 1 at -1 ps) is fitted with a single exponential decay (shown in Fig. S8, left panel), which is assumed to result from transient heating. This exponential reduction is then compared with the purely thermal reduction from the static measurements. The comparison is made by fitting the ZRS intensity as a function of temperature with a third order polynomial in order to produce a relation between ZRS reduction and temperature (Fig. S8, right panel). We then map the reduction of the ZRS intensity as a function of delay onto the static temperature evolution, and from this extract an effective temperature. The saturation of ZRS reduction at long time delays is clearly seen to be below the thermal saturation in the static high temperature measurements.

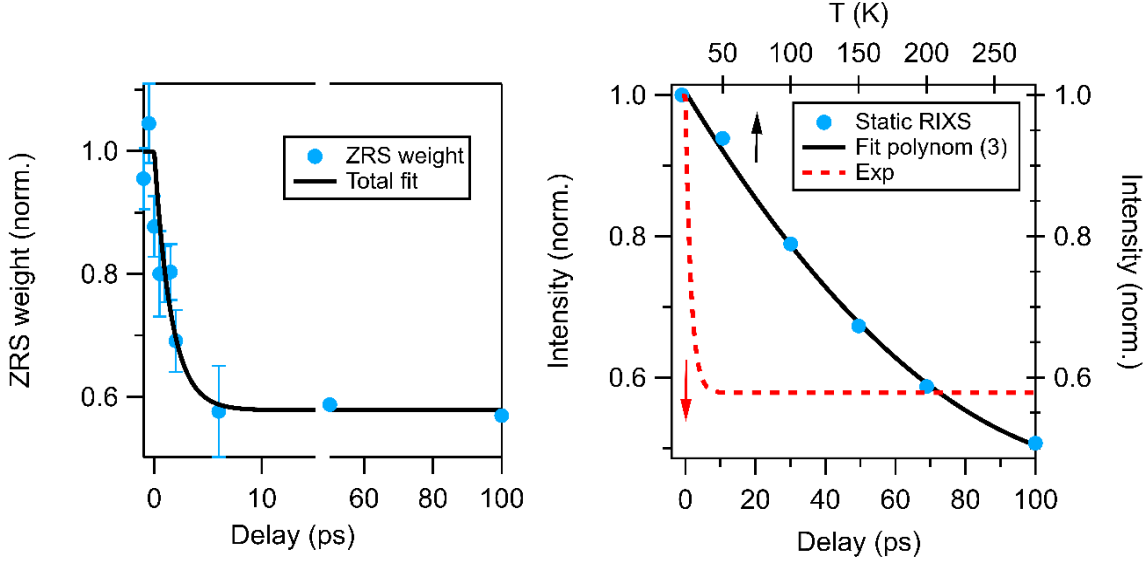

**Figure S 8: Extraction of thermal ZRS dynamics and comparison with static RIXS measurements as a function of temperature.** The reduction in the ZRS intensity as a function of delay is converted into an effective temperature by assuming a one-to-one correspondence with the temperature-dependent reduction obtained from the static RIXS data.

Furthermore, we model this temperature evolution as described below, in order to compare the extracted temperature with that expected, given the heat input from the pump pulse. Using a simple two-temperature model (TTM) allows an estimation of the transient temperature of electrons and lattice based on the experimental pump pulse parameters. The model consists of two coupled differential equations:

$$\frac{\partial T_e}{\partial t} = \frac{P}{C_e}\sqrt{\frac{2.77}{\pi\sigma^2}}e^{\left(-2.77(t-t_0)^2/\sigma^2\right)} - \frac{g}{C_e}(T_e - T_l),$$

$$\frac{\partial T_l}{\partial t} = \frac{g}{C_l}(T_e - T_l).$$

These two equations describe the evolution of the electronic ($T_e$) and lattice ($T_l$) temperatures. The first term in the electronic temperature equation is the source (pump laser pulse), modeled by a normalized Gaussian, which deposits heat into the electronic system leading to an increase of $T_e$. The second term – the coupling term – is proportional to the difference between the electron and lattice temperatures and therefore transports heat from the electronic into the phononic system, which equilibrate after some time depending on the strength of the electron-phonon coupling.

A number of parameters are used in the two equations. $C_e = \gamma T_e$ and $C_l$ are the electron and lattice specific heat capacities ($\gamma$ is the electronic contribution to the specific heat). The electron-phonon coupling is proportional to $g$. The laser parameters are: $P$, the absorbed energy density within the probed region; $\sigma$, the full-width at half maximum (in time) of the laser pulse; and $t_0$, the time zero. The values of these parameters used in the calculations are given in Table S1 along with the relevant references from literature. $P$ and $\sigma$ are determined experimentally, however as discussed in the main text the value of $P$ estimated from the experiment does not well reproduce the observed temperature dynamics. For this reason in the final two-temperature model $P$ is adjusted to match the experimental temperature dynamics. This difference between the measured and modelled $P$ implies a breakdown of the simple two-temperature model and a non-equilibrium thermal distribution between the lattice and magnetic sub-systems. The temperature dependence of $C_l$ is approximated by a polynomial interpolation from the data in Ref.[7] until the Debye temperature of 360 K (Ref.[8]), above which the tabulated value is used. The equations were solved numerically and the results presented in Fig. 3 (e) of the main text.

| Parameter | Value | Source |
|---|---|---|
| $P$ | 83 J cm$^{-3}$ | Fit |
| $\sigma$ | 70 fs | Experiment |
| $\gamma$ | 0.00154 J cm$^{-3}$ K$^{-2}$ | Ref.[9] |
| $C_l$ | 0.55 J g$^{-1}$ K$^{-1}$ (at 400 K) | Refs.[7,9] |
| $g$ | 0.07 | Fit parameter |

**Table S1. Input parameters for the two-temperature model.**

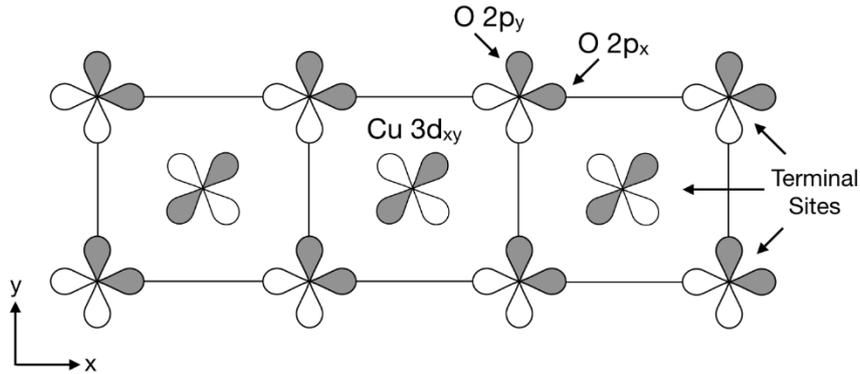

**Figure S9: The Cu$_3$O$_8$ cluster used to model CuGeO$_3$.** The Cu-O-Cu bond angle is 99.1°. While the orbital basis includes three Cu 3d orbitals at each Cu site, only the 3d$_{xy}$ orbitals are shown for clarity. The cluster has open boundary conditions and the site energies of the terminal sites on each end of the chain are increased slightly to account for the missing Cu-Cu and Cu-O interatomic interactions that would present with periodic boundary conditions.

### Section S4: Model Calculations

We calculated the RIXS spectra using small cluster exact diagonalization and the Kramers-Heisenberg formalism. In this case, the Cu chains of CuGeO$_3$ were modeled using a Cu$_3$O$_8$ cluster with open boundary conditions, as shown in Fig. S9. The orbital basis for this cluster included three Cu $3d_\alpha$ orbitals ($\alpha = xy, x^2 - y^2$, and $3z^2 - r^2$) at each copper site and two oxygen $2p_\beta$ orbitals ($\beta = x, y$) at each oxygen site (twenty-five orbitals in total). The model is similar to the one used in Ref.[10] but modified to include two additional Cu 3d orbitals.

The cluster Hamiltonian is given by $H = H_0 + H_d + H_p + H_{pd} + H_{dd}$, where

$$H_0 = \sum_{m,\alpha,\sigma} \varepsilon^d_{m,\alpha} n^d_{m,\alpha,\sigma} + \sum_{l,\beta,\sigma} \varepsilon^p_{l,\beta} n^p_{l,\beta,\sigma} + \sum_{\langle m,l,\alpha,\beta\rangle,\sigma} t^{m,l}_{\alpha,\beta}[d^\dagger_{m,\alpha,\sigma} p_{l,\beta,\sigma} + \text{h.c.}]$$
$$+ \sum_{\langle l,l',\beta,\beta'\rangle,\sigma} t^{l,l'}_{\beta,\beta'}\left[p^\dagger_{l',\beta',\sigma} p_{l,\beta,\sigma} + \text{h.c.}\right]$$

describes noninteracting terms,

$$H_d = \sum_{m,\alpha} U^d_{\alpha,\alpha} n^d_{m,\alpha,\uparrow} n^d_{m,\alpha,\downarrow} + \sum_{\substack{m,\sigma,\sigma'\\\alpha\neq\alpha'}} \frac{U^d_{\alpha,\alpha'}}{2} n^d_{m,\alpha,\sigma} n^d_{m,\alpha',\sigma'} + \sum_{\substack{m,\sigma,\sigma'\\\alpha\neq\alpha'}} \frac{J^d_{\alpha,\alpha'}}{2} d^\dagger_{m,\alpha,\sigma} d^\dagger_{m,\alpha',\sigma'} d_{m,\alpha,\sigma'} d_{m,\alpha',\sigma}$$
$$+ \sum_{\substack{m,\sigma\neq\sigma'\\\alpha\neq\alpha'}} \frac{J^d_{\alpha,\alpha'}}{2} d^\dagger_{m,\alpha,\sigma} d^\dagger_{m,\alpha,\sigma'} d_{m,\alpha',\sigma'} d_{m,\alpha',\sigma}$$

and

$$H_p = \sum_{l,\beta} U_p n^p_{l,\beta,\uparrow} n^p_{l,\beta,\downarrow} + \sum_{\substack{l,\sigma,\sigma'\\\alpha\neq\alpha'}} \frac{U'_p}{2} n^p_{l,\beta,\sigma} n^p_{l,\beta',\sigma'} + \sum_{\substack{l,\sigma,\sigma'\\\alpha\neq\alpha'}} \frac{J_p}{2} p^\dagger_{l,\beta,\sigma} p^\dagger_{l,\beta',\sigma'} p_{l,\beta,\sigma'} p_{l,\beta',\sigma}$$
$$+ \sum_{\substack{l,\sigma\neq\sigma'\\\beta\neq\beta'}} \frac{J_p}{2} p^\dagger_{l,\beta,\sigma} p^\dagger_{l,\beta,\sigma'} p_{l,\beta',\sigma'} p_{l,\beta',\sigma}$$

describes the intra-atomic interactions on the Cu and O atoms, respectively, and

$$H_{dd} = \sum_{\langle m,m',\alpha,\alpha'\rangle,\sigma,\sigma'} \frac{V_{dd}}{2} n^d_{m,\alpha,\sigma} n^d_{m',\alpha',\sigma'}$$

and

$$H_{pd} = \sum_{\substack{\langle m,l,\alpha,\beta\rangle\\\sigma,\sigma'}} [U_{pd} - K_{pd}\delta_{\sigma,\sigma'}] n^d_{m,\alpha,\sigma} n^p_{l,\beta,\sigma'} - K_{pd} \sum_{\substack{\langle m,l,\alpha,\beta\rangle\\\sigma}} \left[d^\dagger_{m,\alpha,\sigma} d_{m,\alpha,\bar\sigma} p^\dagger_{l,\beta,\bar\sigma} p_{l,\beta,\sigma} + \text{h.c.}\right]$$
$$+ K_{pd} \sum_{\langle m,l,\alpha,\beta\rangle} \left[d^\dagger_{m,\alpha,\uparrow} p_{l,\beta,\uparrow} d^\dagger_{m,\alpha,\downarrow} p_{l,\beta,\downarrow} + \text{h.c.}\right]$$

describes nearest-neighbor Cu-O and Cu-Cu interatomic interactions, respectively. Here, $\langle \ldots \rangle$ denotes a sum over nearest neighbor orbitals, $d^\dagger_{m,\alpha,\sigma}$ ($d_{m,\alpha,\sigma}$) creates (annihilates) a spin-$\sigma$ hole in the $3d_\alpha$ orbital of Cu atom $m$, $p^\dagger_{l,\beta,\sigma}$ ($p_{l,\beta,\sigma}$) creates (annihilates) a spin-$\sigma$ hole in the $2p_\beta$ orbital of O atom $l$, and $n^d_{m,\alpha,\sigma} = d^\dagger_{m,\alpha,\sigma} d_{m,\alpha,\sigma}$ and $n^p_{l,\beta,\sigma} = p^\dagger_{l,\beta,\sigma} p_{l,\beta,\sigma}$ are the corresponding number operators.

Throughout this work, we obtain the initial and final states of the RIXS process by diagonalizing the Hamiltonian in the (2 ↑, 1 ↓)-hole sector. The intermediate states are obtained by diagonalizing the problem in the (2 ↑, 0 ↓) and (1 ↑, 1 ↓) sectors, while the third hole is completely localized in the 1s core orbital where it was created. When solving for the intermediate states of the RIXS process, we include the interaction between the valence holes and the O 1s core hole

$$H_{\text{ch}} = U_q \sum_{l,\beta,\sigma,\sigma'} n^p_{l,\beta,\sigma} n^s_{l,\sigma'},$$

where $n_{l,\sigma}^s = s_{l,\sigma}^\dagger s_{l,\sigma}$ is the number operator for the $1s$ orbital at site $l$.

To capture the short-time dynamics of CuGeO$_3$, we introduced a coupling to a coherent phonon oscillation whose displacement is given by

$$X(t) = A\sin(\omega_{\text{mode}}t)\exp\left(-\frac{t}{\tau}\right)\Theta(t).$$

Here, $A$ is the amplitude of the oscillation, $\omega_{\text{mode}}$ is the frequency, $\tau$ is the damping coefficient, and $\Theta(t)$ is the Heaviside step function. The displacement of the oscillations couples linearly and uniformly to the site energy of the Cu $3d_{xy}$ orbitals with $\varepsilon_{xy}^d(t) = \varepsilon_{xy}^d + gX(t)$. This assumption is consistent with long-wavelength oscillation that is expected to be excited at q = 0. (We neglected the coupling to the remaining orbitals for simplicity.) To capture the long-time dynamics, which are dominated by heating, we also introduced an effective temperature parameterized by $T(t) = T_0 + T_1(1 - e^{-t/\tau_t})$. Both of these approximations introduce a time dependence to the Hamiltonian. To evaluate the RIXS intensity, we diagonalized $H(t)$ at each time and used the instantaneous eigenstates and effective temperature $T(t)$ to evaluate the Kramers-Heisenberg formula (see below).

We denote the momentum, energy, and polarization ($\sigma$ or $\pi$) for the incoming and outgoing x-rays as ($\mathbf{k}_{\text{in}}$, $\omega_{\text{in}}$, $\hat{\nu}$) and ($\mathbf{k}_{\text{out}}$, $\omega_{\text{out}}$, $\hat{\mu}$), respectively ($\hbar = 1$). The RIXS intensity at temperature $T$ is then given by the Kramers-Heisenberg formula

$$I_{\mu\nu}(\mathbf{q}, \omega_{\text{in}}, \Omega, T) \propto \frac{1}{Z}\sum_{f,i} g_i e^{\beta E_i}\left|M_{fi}^{\mu\nu}\right|^2 \delta(E_i - E_f + \Omega),$$

where $Z = \sum_i g_i e^{\beta E_i}$ is the partition function, $g_i$ is the degeneracy of the initial state, $\beta = 1/k_b T$ is the inverse temperature, $k_b$ is Boltzmann's constant, $M_{fi}^{\mu\nu}$ is the scattering amplitude between the initial and final states of the RIXS process, and $q = (\mathbf{k}_{\text{in}} - \mathbf{k}_{\text{out}}) \cdot \hat{x}$ and $\Omega = \omega_{\text{in}} - \omega_{\text{out}}$ are the momentum and energy transferred along the chain direction, respectively.

The scattering amplitude is given by

$$M_{fi}^{\mu\nu} = \sum_{m,l} e^{-i\mathbf{q}\cdot\mathbf{R}_l} \frac{\langle f|D_{\mu,l}^\dagger|m\rangle\langle m|D_{\nu,l}|i\rangle}{E_i + \omega_{\text{in}} - E_m + i\Gamma},$$

where $R_l$ is the position of oxygen atom $l$ and the sum over $l$ is taken over all unique oxygen sites in the cluster; $|i\rangle$, $|m\rangle$, and $|f\rangle$ are the initial, intermediate, and final states of the RIXS process with energies $E_i$, $E_m$, and $E_f$; $\Gamma$ is the inverse core-hole lifetime measured in eV, and

$$D_{\nu,l} = \sum_\sigma \left[(\hat{\nu}\cdot\hat{x})p_{l,x,\sigma} s_{l,\sigma}^\dagger + (\hat{\nu}\cdot\hat{y})p_{l,y,\sigma} s_{l,\sigma}^\dagger\right]$$

and

$$D_{\mu,l}^\dagger = \sum_\sigma \left[(\hat{\mu}\cdot\hat{x})s_{l,\sigma}^\dagger p_{l,x,\sigma} + (\hat{\mu}\cdot\hat{y})s_{l,\sigma}^\dagger p_{l,y,\sigma}\right]$$

are the dipole transition operators. For the experimental scattering geometry, $\hat{\sigma}\cdot\hat{x} = 0$, $\hat{\sigma}\cdot\hat{y} = \cos(56°)$, $\hat{\pi}\cdot\hat{x} = \cos(45°)$, and $\hat{\pi}\cdot\hat{y} = 0$. Since the outgoing polarization is not measured during the experiment, the total RIXS intensity is given by $I_\pi(\mathbf{q}, \omega_{\text{in}}, \Omega, T) = \sum_\mu I_{\mu\pi}(\mathbf{q}, \omega_{\text{in}}, \Omega, T)$.

All parameters are listed in units of eV unless otherwise stated. The nearest-neighbor Cu-O hopping integrals $t_{\alpha,\beta}^{m,l}$ are determined using the Slater-Koster parameterization with $(pd\sigma) = 1.6$ and $(pd\pi) = -0.72$. Note that the Cu-O-Cu bond angle in CuGeO$_3$ is 99.1°,[10] which must be accounted for when computing the direction cosines. The nearest-neighbor O-O hopping integrals $t_{\beta,\beta'}^{l,l'}$ are parameterized using $(pp\sigma) = 1$ and $(pp\sigma) = 0.96$ for hopping along x- and y-directions, respectively, which reflects a slight difference in the O-O bond length along these two directions[10]. In both directions, $(pp\pi) = -(pp\sigma)/4$.

The onsite energies are $\varepsilon_{xy}^d = 0$ (0.5), $\varepsilon_{x^2-y^2}^d = 1.2$ (1.7), $\varepsilon_{3z^2-r^2}^d = 1.6$ (2.1), $\varepsilon_x^p = 5.6$ (6.58), and $\varepsilon_y^p = 5.25$ (6.23). The values listed in brackets are for the terminal sites of the chain, as indicated in Fig. S9. The interatomic Hubbard and Hund's interactions for the Cu sites are listed in Table S2. The remaining interaction parameters are $U_p = 4.1, J_p = 0.6, U_p' = U_p - 2J_p, V_{dd} = 0.5, U_{pd} = 1.2, K_{pd} = 0.11, U_q = 5$. The parameters related to the e-ph coupling are $g = 4$ eV/Å, $A = 0.07$ Å, $\tau = 0.5$ ps, and $\omega_{\text{mode}} = 0.0031$ fs$^{-1}$. (The values of these parameters were adjusted to match the experimental data.) The parameters related to the effective temperature are $T_0 = 21.3$ K, $T_1 = 181.2$ K, and $\tau_t = 1.949$ ps, which mimic the behavior estimated from the heating analysis presented in the main text. The remaining parameters relevant for computing the RIXS intensity are $\Gamma = 150$ meV and $\omega_{\text{in}} = 0.5$ eV, which corresponds to the peak in the XAS spectra for our model (see Ref.[6]).

| $U_{\alpha,\alpha'}^d$ ($J_{\alpha,\alpha'}^d$) | $\alpha = 3d_{xy}$ | $\alpha = 3d_{x^2-y^2}$ | $\alpha = 3d_{3z^2-r^2}$ |
|---|---|---|---|
| $\alpha' = 3d_{xy}$ | 10 (n.a.) | 7.8 (0.35) | 5.8 (1.35) |
| $\alpha' = 3d_{x^2-y^2}$ | 7.8 (0.35) | 10 (n.a.) | 5.8 (1.35) |
| $\alpha' = 3d_{3z^2-r^2}$ | 5.8 (1.35) | 5.8 (1.35) | 10 (n.a.) |

**Table S2: The values of the interatomic Cu Hubbard $U_{\alpha,\alpha'}^d$ and Hund's $J_{\alpha,\alpha'}^d$ parameters used in this work**, given in units of eV. The values listed in parentheses correspond to the Hund's parameters.

# Supplementary References

1. Schlotter, W. F. *et al.* The soft x-ray instrument for materials studies at the linac coherent light source x-ray free-electron laser. *Rev. Sci. Instrum.* **83**, 43107 (2012).

2. http://henke.lbl.gov/optical_constants/.

3. Giannetti, C. *et al.* Disentangling thermal and nonthermal excited states in a charge-transfer insulator by time- and frequency-resolved pump-probe spectroscopy. *Phys. Rev. B* **80**, 235129 (2009).

4. Pagliara, S., Parmigiani, F., Galinetto, P., Revcolevschi, A. & Samoggia, G. Role of the Zhang-Rice-like exciton in optical absorption spectra of CuGeO3 and CuGe1-xSixO3 single crystals. *Phys. Rev. B* **66**, 24518 (2002).

5. Dhalenne, G., Revcolevschi, A., Rouchaud, J. C. & Federoff, M. FLOATING ZONE CRYSTAL GROWTH OF PURE AND Si- OR Zn-SUBSTITUTED COPPER GERMANATE CuGeO3. *Mater. Res. Bull.* **32**, 939–945 (1997).

6. Monney, C. *et al.* Determining the Short-Range Spin Correlations in the Spin-Chain

Li2CuO2 and CuGeO3 Compounds Using Resonant Inelastic X-Ray Scattering. *Phys. Rev. Lett.* **110**, 87403 (2013).

7. Weiden, M. *et al.* Thermodynamic properties of the spin-Peierls transition in CuGeO3. *Zeitschrift für Phys. B Condens. Matter* **98**, 167–169 (1995).

8. Kuo, Y.-K., Figueroa, E. & Brill, J. W. Mean-field specific heat of CuGeO3. *Solid State Commun.* **94**, 385–389 (1995).

9. Lasjaunias, J. C. *et al.* Heat capacity of CuGeO3: sensitivity to crystalline quality. *Solid State Commun.* **101**, 677–680 (1997).

10. Mizuno, Y. *et al.* Electronic states and magnetic properties of edge-sharing Cu-O chains. *Phys. Rev. B* **57**, 5326–5335 (1998).